\documentclass[usegraphicx,usenatbib]{mn2e} 

\usepackage{url}
\usepackage[breaklinks=true]{hyperref}
\usepackage{twoopt}

\usepackage{natbib}
\bibpunct{(}{)}{;}{a}{}{,} 

\usepackage[utf8x]{inputenc}
\usepackage[normalem]{ulem}
\usepackage{siunitx}
\usepackage{amsmath, amssymb}
\usepackage[T1]{fontenc}
\usepackage{aecompl}

\usepackage[farskip=0pt]{subfig}

\usepackage{multirow,bigstrut,ctable}
\usepackage[normalem]{ulem}
\usepackage{rotating}
\usepackage[varg]{txfonts} 
\usepackage{threeparttable}

\def \deg         {\text{$^{\circ}$}}
\def \arcmin      {\text{$^\prime$}}
\def \arcsec      {\text{$^{\prime\prime}$}}
\def \hour        {$^{\mathrm{h}}$}
\def \min         {$^{\mathrm{m}}$}
\def \sec         {$^{\mathrm{s}}$}
\def \mjybeam     {mJy\,beam$^{-1}$}
\def \mujybeam    {$\rm \mu$Jy\,beam$^{-1}$}

\newcommand{\Msun}{\text{$\rm M_\odot$}}

\newcommand{\hms}[3]{{#1}\hour{#2}\min{#3}\sec}
\newcommand{\dms}[3]{{#1}\deg{#2}\arcmin{#3}\arcsec}
\newcommand{\beam}[2]{{#1}\arcsec$\times${#2}\arcsec}

\newcommand{\arcmind}[2]{$#1'\,\hspace{-1.5mm}.\hspace{.1mm}#2$}

\def \target       {Abell 1033}

\title[Abell 1033: birth of a radio phoenix]
      {Abell 1033: birth of a radio phoenix}

\author[F.~de~Gasperin et~al.]{F. de Gasperin$^{1}$, G.~A.~Ogrean$^{1,2}$, R. J. van Weeren$^{2}$, W. A. Dawson$^{3}$, M. Br\"uggen$^{1}$,\newauthor 
A. Bonafede$^{1}$, A. Simionescu$^{4}$
\\
$^{1}$ Universit\"at Hamburg, Hamburger Sternwarte, Gojenbergsweg 112, D-21029, Hamburg, Germany\\
$^{2}$ Harvard-Smithsonian Center for Astrophysics, 60 Garden Street, Cambridge, MA 02138, USA\\
$^{3}$ Lawrence Livermore National Lab, 7000 East Avenue, Livermore, CA 94550, USA\\
$^{4}$ Japan Aerospace Exploration Agency, 3-1-1 Yoshinodai, Sagamihara, Kanagawa 229-8510, Japan\\
}
\begin{document}

\date{}
\pagerange{\pageref{firstpage}--\pageref{lastpage}} \pubyear{2014}
\maketitle

\label{firstpage}

\begin{abstract}
Extended steep-spectrum radio emission in a galaxy cluster is usually associated with a recent merger. However, given the complex scenario of galaxy cluster mergers, many of the discovered sources hardly fit into the strict boundaries of a precise taxonomy. This is especially true for radio phoenixes that do not have very well defined observational criteria. Radio phoenixes are aged radio galaxy lobes whose emission is reactivated by compression or other mechanisms. Here, we present the detection of a radio phoenix close to the moment of its formation. The source is located in Abell 1033, a peculiar galaxy cluster which underwent a recent merger. To support our claim, we present unpublished Westerbork Synthesis Radio Telescope and \textit{Chandra} observations together with archival data from the Very Large Array and the Sloan Digital Sky Survey. We discover the presence of two sub-clusters displaced along the N-S direction. The two sub-clusters probably underwent a recent merger which is the cause of a moderately perturbed X-ray brightness distribution. A steep-spectrum extended radio source very close to an AGN is proposed to be a newly born radio phoenix: the AGN lobes have been displaced/compressed by shocks formed during the merger event. This scenario explains the source location, morphology, spectral index, and brightness. Finally, we show evidence of a density discontinuity close to the radio phoenix and discuss the consequences of its presence.
\end{abstract}

\begin{keywords}
  galaxies: clusters: individual: \target{} -- large-scale structure of Universe -- radio continuum: general -- X-rays: galaxies: clusters
\end{keywords}

\section{Introduction}
\label{sec:introduction}

Galaxy clusters form via a hierarchical sequence of major and minor mergers. A merger event dissipates up to $10^{63}-10^{64}$~ergs into heating the gas but also through large-scale intra-cluster medium (ICM) motions such as shocks and turbulence. These processes are related to the formation of non-thermal components in the ICM: cosmic rays (CR) and magnetic fields \citep[e.g.][]{Brunetti2014}. Relativistic particles diffusing into magnetic fields emit synchrotron radiation. Therefore, non-thermal components in galaxy clusters are most evident once their radio emission is traced. Current observations strongly support this idea of a tight connection between cluster-scale synchrotron emission and mergers, finding indeed these radio sources always hosted in dynamically disturbed systems \citep{Cassano2010,Cassano2013}. 

Radio relics and radio halos are two types of radio sources associated with non-thermal cluster components. \textit{Radio relics} (or gischt) are located in cluster outskirts. They are long (1--2 Mpc) and thin ($\sim100$ kpc) sources. They are polarized and are likely signatures of electrons accelerated by large-scale accretion shocks through, for example, diffusive shock acceleration \citep{Drury1983,Blandford1987,Jones1991}. \textit{Radio halos} are diffuse, unpolarized, centrally located sources. They may be caused by large turbulent regions formed during cluster collisions and can reaccelerate populations of long-lived synchrotron-dark CR electrons to radio bright energies \citep{Donnert2013}. These populations are generated by AGN activity, shocks or supernovae in galaxies \citep{Brunetti2014}.

A third, less studied, type of emission in merging clusters are the so-called \textit{radio phoenixes} \citep{Slee2001,Ennslin2002,Kempner2004,VanWeeren2009b}. When shocks from a cluster merger pass through an old AGN lobe, they compress the fossil radio plasma. Because the density in the fossil plasma is much lower than the surrounding ICM, these shocks may become subsonic compression waves inside the plasma, as shown by \cite{Ensslin2001}. This compression is able to re-energizing the electrons in the fossil plasma so they shine again at radio wavelengths.

Here, we present the detection of a radio phoenix located in \target{}, a peculiar galaxy cluster which underwent a recent merger. In Sec.~\ref{sec:observation} we describe the radio, optical and X-ray observations of the galaxy cluster and related data reduction. In Sec.~\ref{sec:cluster} we present the results of these observations and we make a comprehensive description of \target{}. Discussion and conclusions are in Sec.~\ref{sec:discussion} and Sec.~\ref{sec:conclusions}, respectively. Throughout this paper we assume a $\Lambda$ cold dark matter ($\Lambda$CDM) cosmology with $H_0 = 73$ km s$^{-1}$ Mpc$^{-1}$, $\Omega_m$ = 0.27 and $\Omega_\Lambda$ = 0.73. At the redshift of \target{}'s brightest cluster galaxy \citep[$z=0.1259 \pm 0.0006$;][]{Allen1992} 1\arcsec{} corresponds to $2.171$ kpc. All images are in the J2000 coordinate system.

\section{Observations}
\label{sec:observation}

The study of \target{} has been carried our using new Westerbork Synthesis Radio Telescope (WSRT, 1.4 GHz) observations and archival \textit{Chandra}, Very Large Array (VLA, 1.4 GHz, B-configuration), and Sloan Digital Sky Survey (SDSS) data.

\subsection{WSRT}
\label{sec:WSRT}

\target{} was observed with the WSRT\footnote{WSRT: \url{http://www.astron.nl}} on June 6, 2014, for 12 h with the default 21~cm setup. Only 9 antennas participated in the observation because of ongoing upgrades. A total bandwidth of 160~MHz was recorded, spread over eight spectral windows of 20~MHz in bandwidth, each having 64 channels. All four linear polarization products were recorded. The calibrators 3C147 and 3C286 were observed at the start and end of the main observing run, respectively.

We calibrated the data with CASA\footnote{\label{fn:casa}CASA version 4.2.1: \url{http://casa.nrao.edu}}. The first step in the data reduction consisted of the removal of time ranges affected by shadowing. We then flagged radio frequency interference using the AOFlagger \citep{Offringa2012}. Before flagging, data were corrected for the bandpass response using 3C147. This prevents flagging of data due to the bandpass rolloff at the edges of the spectral windows. After flagging the data with AOFlagger, we again determined the bandpass response using 3C147. We then proceeded with standard gain calibration using the \cite{Perley2012} flux scale. The channel dependent polarization leakage terms were calibrated using 3C147, assuming the source is unpolarized. The channel dependent polarization angles were set using 3C286. As a last step all calibration solutions were transferred to the target field.

\begin{figure*}
\centering
\subfloat[WSRT 1.4 GHz]{\includegraphics[width=.49\textwidth]{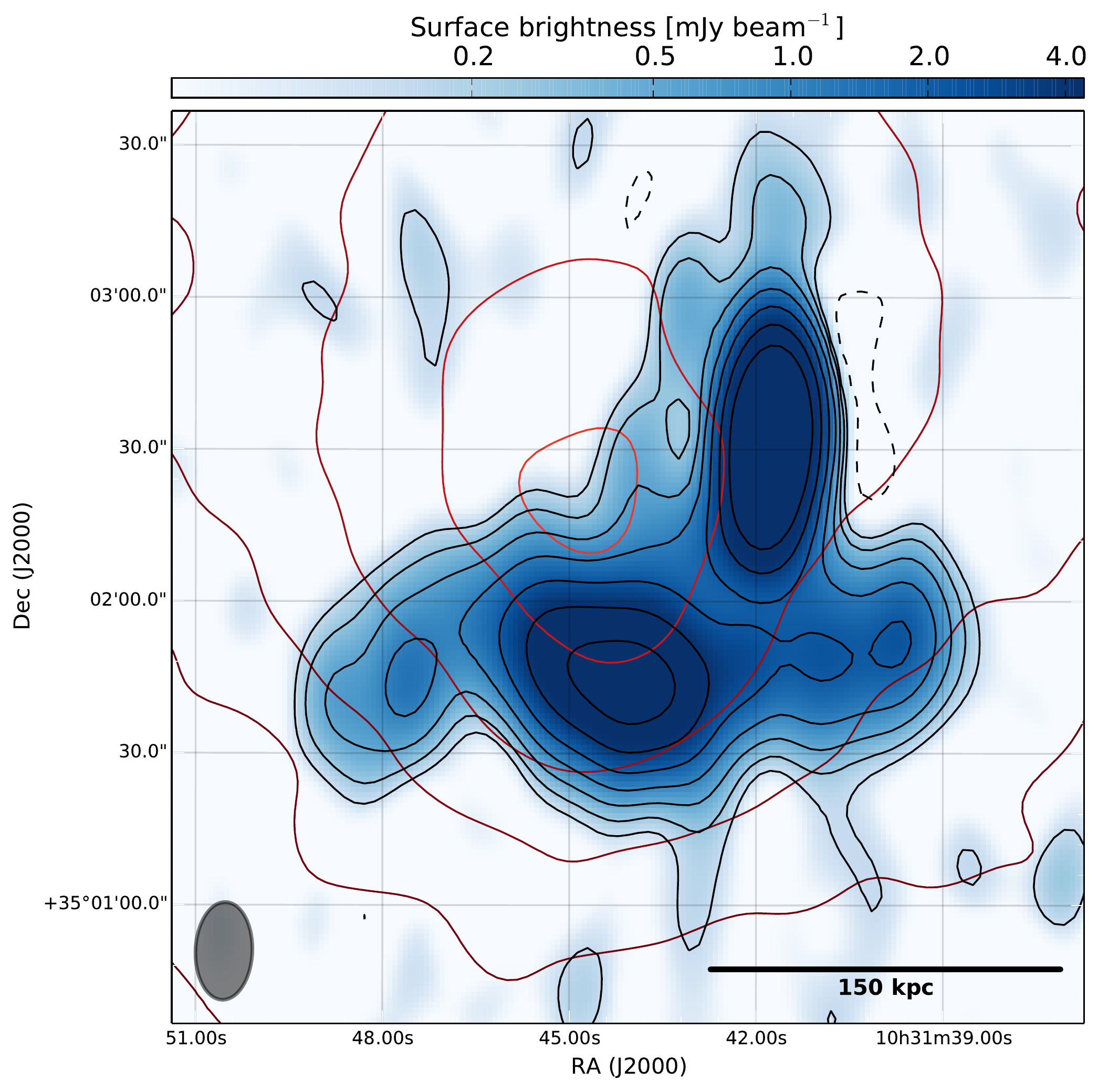}\label{fig:radio-wsrt}}
\subfloat[VLA 1.4 GHz]{\includegraphics[width=.49\textwidth]{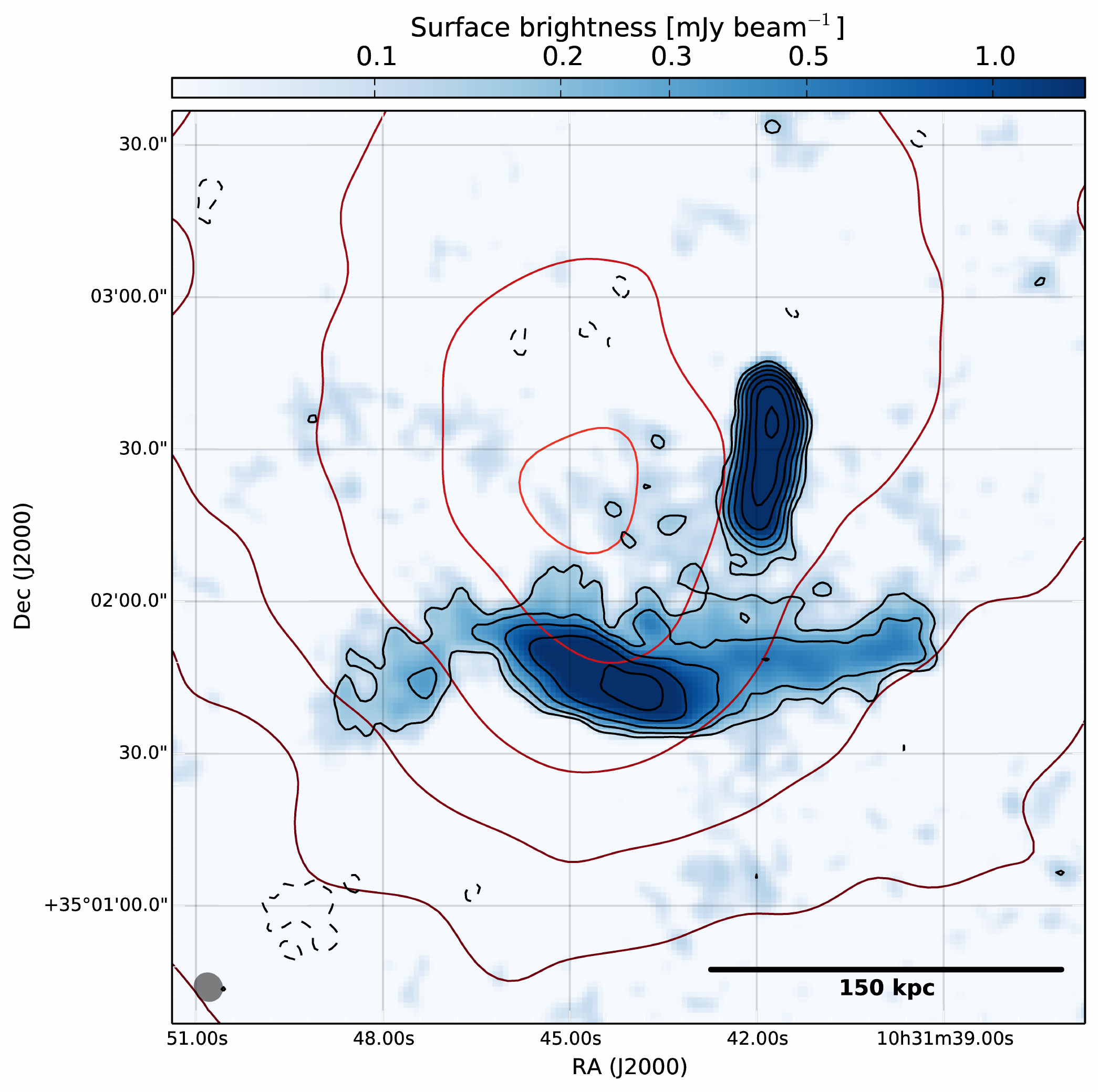}\label{fig:radio-vla}}
\caption{\textit{Left}: WSRT image at 1.4 GHz, contours are at $\left[-1, 1, 2, 4, 8, 16, 32, 64\right] \times 3 \sigma$, with $\sigma=44$~\mujybeam (beam = \beam{19}{11}). \textit{Right}: VLA image at 1.4 GHz, contours are at $\left[-1, 1, 2, 4, 8, 16, 32, 64\right] \times 3 \sigma$, with $\sigma=47$~\mujybeam (beam = \beam{5}{5}). In both images red contours trace the X-ray emission as described in Fig.~\ref{fig:Chandra}.}\label{fig:radio} 
\end{figure*}

For the target field, we carried out several cycles of phase-only self-calibration on 3~min timescales. This was followed by a final round of amplitude and phase self-calibration on 20 min timescales (pre-applying the previous phase-only self-calibration on shorter timescales). The data was imaged taking the spectral index into account during the deconvolution \cite[i.e., {\tt nterms=2}][]{Rau2011}. The final image (see Fig.~\ref{fig:radio-wsrt}) uses uniform weighting and it is corrected for the primary beam attenuation. The resolution is \beam{19}{11} and the rms noise is $\sigma=44$~\mujybeam. A low-resolution image has been obtained after $uv$-subtracting the high-resolution clean components and tapering the data. This image has a resolution of \beam{41}{34}, an rms noise of $100$~\mujybeam and is shown in Fig.~\ref{fig:wsrt-ext}.

\subsection{VLA}
\label{sec:VLA}

Searching the VLA\footnote{VLA: \url{http://www.vla.nrao.edu}} archive we found an observation at 1425 MHz taken on 18th April 1993 with the ``old'' VLA in B-configuration (project code AA150). The target was observed for a total of 30 minutes using NVSS J112553+261020 as phase calibrator and 3C\,286 as flux calibrator. Fluxes have been scaled to match the \cite{Perley2012} flux scale. Data were taken in two spectral windows centred on 1.39 GHz and 1.46 GHz. The bandwidth was of 50 MHz in both cases. After a standard calibration made with the CASA\textsuperscript{\ref{fn:casa}} package, phases were self-calibrated and the visibilities were imaged using the CLEAN task with a Briggs weighting of +1 to enhance the extended emission. The final image (Fig.~\ref{fig:radio-vla}) has a resolution of 5\arcsec{} and an rms noise of 47~\mujybeam.

\subsection{Optical Imaging and Spectroscopy}
\label{sec:Optical}
We use photometry and spectroscopic/photometric redshifts from the $10^{\rm th}$ release of the Sloan Digital Sky Survey \citep{Ahn2014} to investigate the dynamics of \target{}. We limit our optical investigation to a $20\arcmin\times20\arcmin$ ($2.7\times2.7$~Mpc$^2$) field-of-view centred on RA = \hms{10}{31}{42} and Dec = \dms{+35}{03}{36}. The brightest cluster galaxy (BCG) does not have an SDSS spectroscopic redshift, however it was observed by \citet{Allen1992} which found a redshift of $0.1259\pm0.0006$. This is the only galaxy within the $20\arcmin\times20\arcmin$ field-of-view with a redshift in the NASA/IPAC Extragalactic Database (NED)\footnote{NED: \url{http://ned.ipac.caltech.edu}} that does not have an SDSS spectroscopic redshift. We include this redshift with the SDSS spectroscopic redshift catalogue for our redshift analysis. In total there are 51 spectroscopic redshifts in the field-of-view, 33 of which are within $\pm3000$\,km\,s$^{-1}$ of the cluster mean spectroscopic redshift ($z=0.1230\pm0.0005$). However, unless specified, in what follows the redshift of the cluster is assumed to be the one of the BCG ($z = 0.1259 \pm 0.0006$). We made this choice since there appears to be a bimodal redshift distribution (see Sec.~\ref{sec:redshift_analysis}) and peak of the X-ray emission and radio phoenix are both close to the BCG in projection.

\subsection{Chandra}
\label{sec:Chandra}

Abell 1033 was observed with \emph{Chandra} for 30~ks on 2013 February~19, and for 34~ks on 2013 February~21 (PI: F.~Gastaldello, ObsIDs 15084 and 15614). The two datasets were retrieved from the \emph{Chandra} Data Archive, and analysed using {\sc ciao} v4.5.9 and the \emph{Chandra} Calibration Database (CALDB) v4.5.9 released on 2013 November 19. Both observations were taken in VFAINT mode, with the ACIS-I and ACIS-S3 CCDs on. The level 1 event files were reprocessed and screened for potential bad events associated with cosmic rays (CR). The new level 2 event files were afterwards screened for background flares using the \textsc{lc\_clean} script. Point sources were identified and removed from the ACIS-S3 CCD, and a good time intervals file was created from the source-free ACIS-S3 event file by calculating the mean $2.5-7$~keV event rate from time bins within $3\sigma$ of the global ACIS-S3 mean in the same energy band, and by then excluding all the time bins in which the event rate deviates from the calculated mean by a factor larger than $1.2$ \citep{Markevitch2002}. The good time intervals defined by this screening were used to remove periods of strong background flares from all the CCDs that were on during the observations. After the flare-filtering, the two observations had clean exposure times of 22.8 and 23.6 ks.

\begin{figure}
\centering
\includegraphics[width=0.43\textwidth]{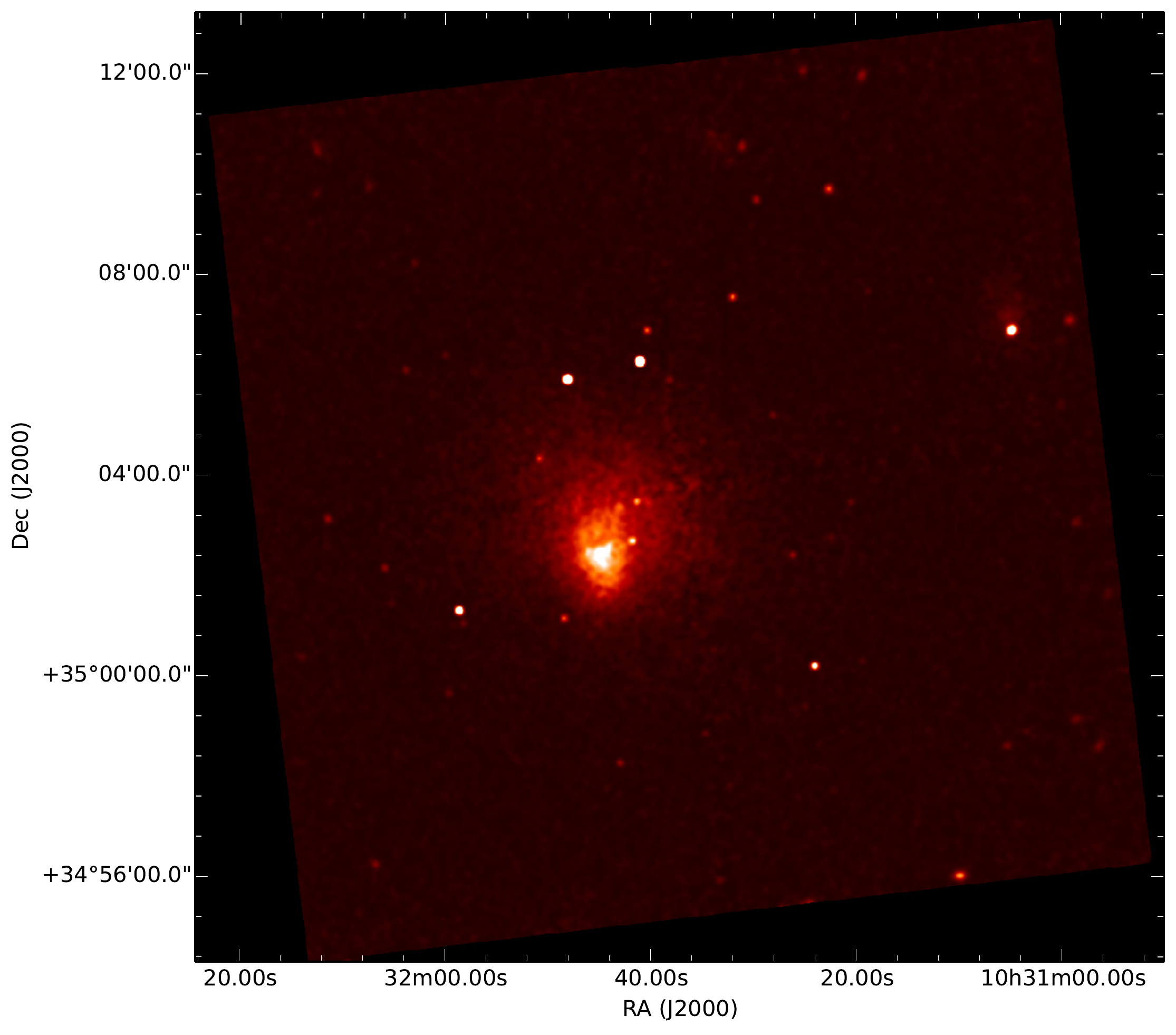}\\
\includegraphics[width=0.43\textwidth]{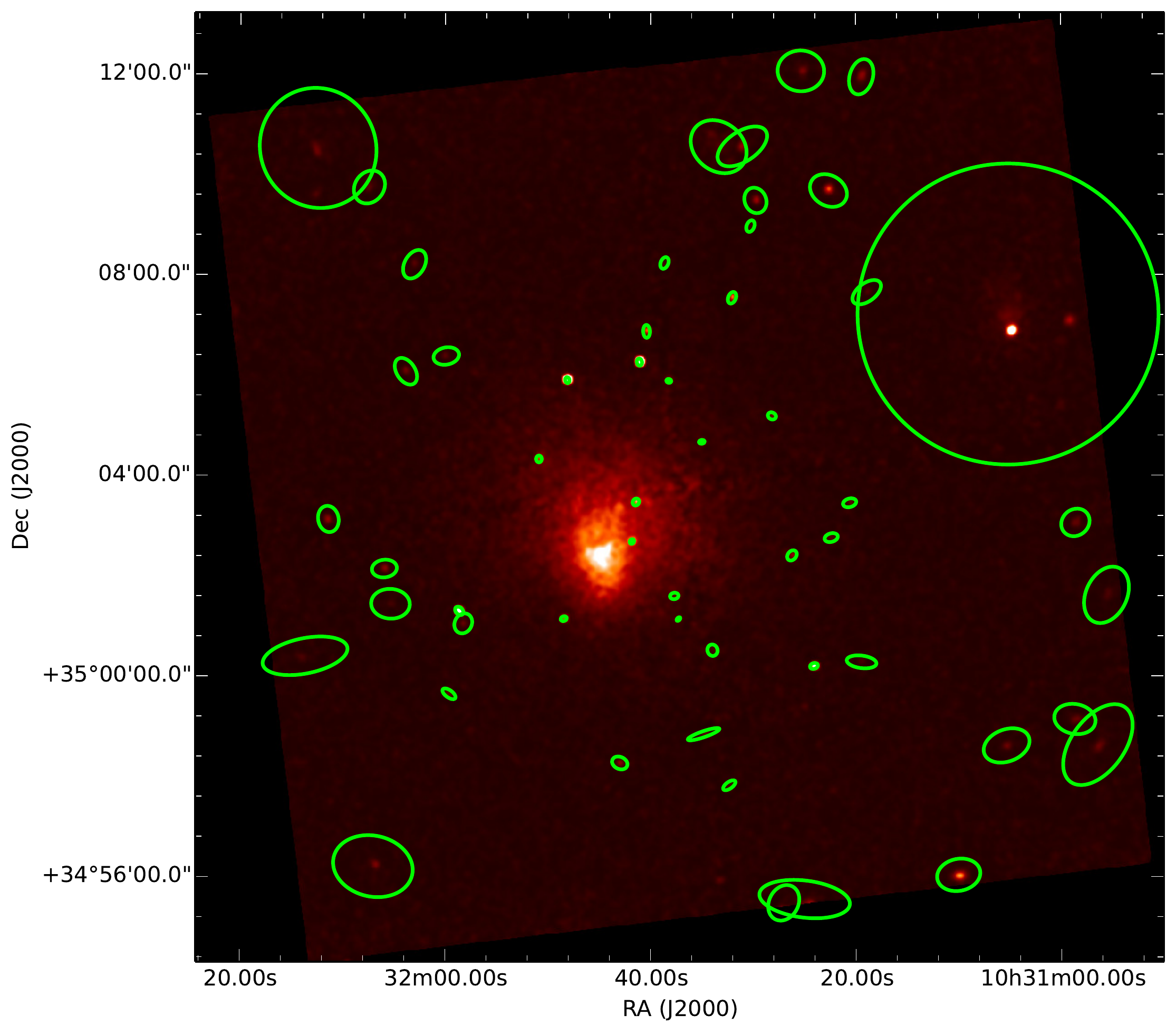}\\
\includegraphics[width=0.43\textwidth]{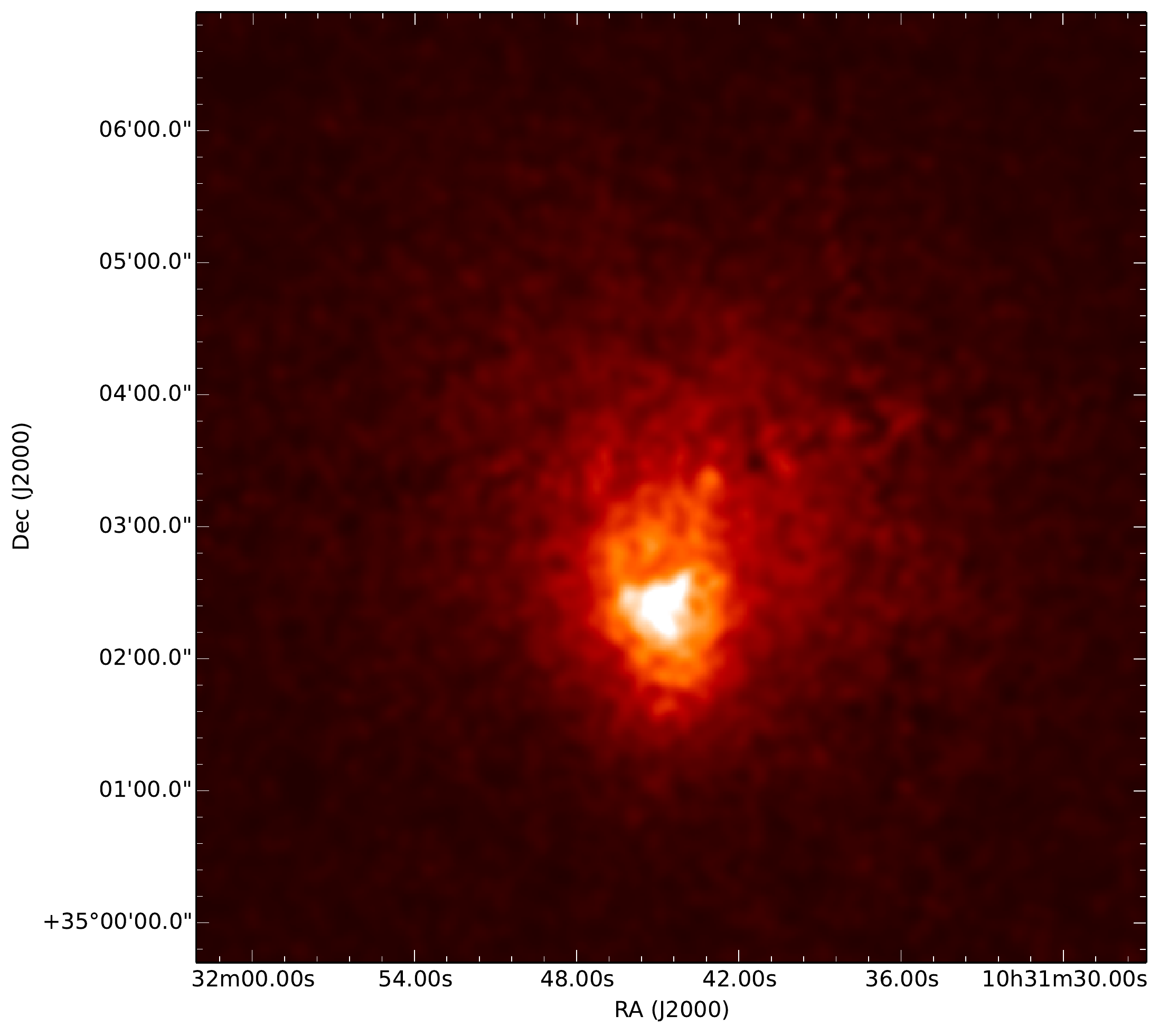}
\caption{\emph{Left}: \emph{Chandra} surface brightness map, in the energy band $0.5-7$~keV. The image was binned by 2, exposure-corrected, vignetting-corrected, and smoothed with a two-dimensional Gaussian kernel of $\sigma=3$~pixels ($1$~pixel $\approx$ 1~arcsec). \emph{Centre}: the image shows point sources detected by \emph{wavdetect} and confirmed visually, as well as two additional point sources missed by \emph{wavdetect} and a circular region of radius $3$~arcmin located west of the cluster, centred approximately on the position of three galaxy cluster candidates listed in the NED database. \emph{Right}: image with the point sources subtracted zoomed in on the main cluster emission.}
\label{fig:chandra-ptsources}
\end{figure}

Count images were created in the energy band $0.5-7$~keV for each of the two Abell 1033 datasets, and binned by a factor of 2 (1 pixel $\approx$ $1^{\prime\prime}$). We also generated corresponding point spread function (PSF) maps with an encircled counts fraction of $90\%$, using an APEC spectral model with a temperature $T=6.5$~keV, a metallicity $Z=0.25$ solar, and a redshift $z=0.1259$; as discussed in Section \ref{sec:globalXray}, these parameters provide a relatively accurate description of the global ICM parameters. Considering the same spectral model and energy band, exposure maps were generated for each observation. Using the exposure maps and the PSF maps of the two datasets, we created a common exposure map-weighted PSF map. We also created summed exposure and count images, which, together with the exposure map-weighted PSF map, were used to detect point sources in the Abell~1033 field of view with the \textsc{ciao} task \emph{wavdetect}. Point sources were detected using wavelet scales of $1, 2, 4, 8, 16$ pixels and excluded in elliptical regions scaled by a factor of $5\sigma$. All point sources were confirmed visually and excluded from the data, together with two additional point sources missed by \emph{wavdetect}. West of Abell~1033, near the edge of the Chandra FoV, there are three galaxy cluster candidates listed in the NED database: GMBCG J157.77147+35.12026 \citep[z=0.3530,][]{Hao2010}, WHL J103103.4+350623 \citep[$z=0.3312$][]{Wen2010}, and NSCS J103107+350552 \citep[$z=0.1600$][]{Lopes2004}. The locations of these cluster candidates roughly coincide, and they are associated with an X-ray bright AGN at RA = \hms{10}{31}{4.8} and DEC = \dms{35}{06}{50.6}. While X-ray emission from another galaxy cluster does not appear to contaminate the Abell~1033 FoV, to be conservative we excluded from the analysis a circular region with a radius of 3~arcmin around the bright AGN. In Fig.~\ref{fig:chandra-ptsources} we show the exposure-corrected, vignetting-corrected \emph{Chandra} map, with the excluded regions overlaid.

Background event files associated with the two observations were created from Group ``E'' stowed background event files (effective since 2005-09-01T00:00:00) that were reprojected to match the observations, and normalized such that their $10-12$~keV count rates match the $10-12$~keV count rates recorded in the corresponding Abell 1033 datasets. 

The sky foreground and background emission was modelled in a region at the edge of the ACIS-I FoV, assuming that it is described by emission from the Local Hot Bubble (LHB), the Galactic Halo (GH), and the Cosmic X-ray Background (CXB). The foreground components (i.e., the LHB and the GH) were modelled with APEC components, while the CXB was modelled with a power-law. The sky background spectra extracted from ObsIDs 15084 and 15614 were fitted in parallel, leaving the temperature and normalizations of all components free in the fit, but linked between the two spectra. The metallicities of the LHB and GH were fixed to the solar values \citep[assuming][abundances]{angr1989}, the redshifts were fixed to zero, and the power-law index of the CXB was fixed to 1.41 \citep{DeLuca2004}. The hydrogen column density was set to the Galactic value of $1.64\times 10^{20}$~atoms~cm$^{-2}$, as measured in the direction of the cluster from the Leiden/Argentine/Bonn Survey of Galactic H\textsc{i} \citep{Kalberla2005}. To constrain the low temperature LHB component, we fitted simultaneously with the sky background spectra a \emph{ROSAT} All-Sky Survey (RASS) background spectrum\footnote{\url{http://heasarc.gsfc.nasa.gov/cgi-bin/Tools/xraybg/xraybg.pl}} extracted from a circular region with a radius of 1 degree around the position of the cluster centre. The CXB normalization of the RASS spectrum was fixed to $8.85\times 10^{-7}$ photons~keV$^{-1}$~cm$^{-2}$~s$^{-1}$~arcmin$^{-2}$, measured at 1~keV \citep{Moretti2003}.


The spectral fits to the sky background are improved by the addition of an extra APEC component with a temperature of $\sim 1$~keV, similar to the Hot Foreground \citep[HF;][]{Yoshino2009} components detected towards, e.g., the Perseus Cluster \citep{Simionescu2011} and CIZA~J2242.8+5301~\citep{Ogrean2013b}. However, unlike these other clusters, Abell~1033 is seen at a higher Galactic latitude. We performed an F-test to evaluate the significance of the HF component in the direction of Abell~1033, and found a F-test probability $p=1.8\%$ (meaning a better fit with the HF component). We also tried fitting the sky background spectra assuming that the presumable HF emission is instead residual emission from the cluster; however, while this fit had an F-test probability $p=0.69\%$ when compared to the fit without a HF/ICM component, we were able to set only an upper limit on the CXB normalization. Abell 1033 has a virial radius $R_{200}=1.03$~Mpc \citep{Rines2013}. The background region is extracted from radii larger than $\sim 0.96$~Mpc from the cluster centre. ICM emission at $\sim R_{200}$ is not typically detected in $\sim 45$~ks \emph{Chandra} observations, and therefore we consider it more likely that foreground emission is responsible for the hot thermal component detected in the sky background spectra.

The fits were done in the energy band $0.5-7$~keV, using \textsc{Xspec} v12.8.2. The two sets of best-fitting parameters -- without and with a HF component -- are summarized in Table \ref{tab:skybkg}. 

\begin{table}
\centering
\begin{threeparttable}
\begin{tabular}{lccc}
\multicolumn{4}{c}{LHB + GH + CXB} \\
\hline
	 	& 		$T$\tnote{a}		& 		$\Gamma$ 	& 		$\mathcal{N}$\tnote{b} 	\\
\hline
      LHB	&       $0.10\pm 0.002$		&			--		&	$(1.81\pm 0.06) \times 10^{-6}$		\\
      GH	&	$0.25_{-0.03}^{+0.04}$	&			--		&	$5.81_{-1.36}^{+1.34}\times 10^{-7}$	\\
      CXB	&			--			&	$1.41^{\dagger}$	&	$(7.13\pm 0.52)\times 10^{-7}$			\\
\hline
\vspace{0.2cm}\\
\multicolumn{4}{c}{LHB + GH + CXB + HF} \\
\hline
	 	& 		$T$\tnote{a}		& 		$\Gamma$ 	& 		$\mathcal{N}$\tnote{b} 	\\
\hline
      LHB	&       $0.10_{-0.03}^{+0.005}$	&			--		&	$1.63_{-0.60}^{+0.15} \times 10^{-6}$	\\
      GH	&	$0.16_{-0.05}^{+0.04}$	&			--		&	$9.72_{-3.70}^{+27.57}\times 10^{-7}$	\\
      CXB	&			--			&	$1.41^{\dagger}$	&	$(6.64\pm 0.56)\times 10^{-7}$			\\
      HF	&	$1.06\pm 0.22$		&			--		&	$1.35_{-0.45}^{+0.50}\times 10^{-7}$	\\
\hline
\end{tabular}
\begin{tablenotes}
	\item[a] temperature, in units of keV;
	\item[b] spectral normalizations, in units of photons~keV$^{-1}$~cm$^{-2}$~s$^{-1}$~arcmin$^{-2}$ measured at 1~keV for the CXB, and in units of cm$^{-5}$~arcmin$^{-2}$ for the LHB, GH, HF, and ICM;
	\item[$\dagger$] fixed parameter.
\end{tablenotes}
\end{threeparttable}
\caption{Best-fitting sky background parameters calculated from the two \emph{Chandra} ObsIDs using different spectral components -- first with LHB, GH, and CXB components, and second with an additional HF component.}
\label{tab:skybkg}
\end{table}

\section{\target}
\label{sec:cluster}
The galaxy cluster \target{} is at $z\sim0.1259$ and has an SZ estimated mass of $M_{500} = (3.4\pm0.4) \times 10^{14}$~\Msun \citep{PlanckCollaboration2013}. In this section we will present the analysis of the cluster’s galaxy distribution in projection and redshift. We will also study the radio environment and the X-ray properties of the cluster.

\subsection{Optical properties}
\subsubsection{Projected Galaxy Distribution}
\label{sec:galdist}
We use SDSS photometry and photometric redshifts to select likely cluster members in order to study the projected galaxy distribution of the cluster. We use the SDSS \emph{type} flag to disambiguate stars and galaxies in the field. We select all galaxies with a photometric redshift within $\pm0.103\left(1+z_\mathrm{cluster}\right)$, which is the root-mean-square of the photometric redshift uncertainty based on the SDSS faint spectroscopic calibration sample\footnote{\url{http://www.sdss3.org/dr10/algorithms/photo-z.php}}. As noted in \S\ref{sec:Optical}, we find the mean cluster redshift $z_\mathrm{cluster}=0.123$. Since the photometric redshifts become unreliable for faint galaxies, due to photometric uncertainty and lack of spectroscopic training galaxies, we apply a magnitude cut of $r<21$ to our cluster sample. Examining the $u-r$ color-magnitude distribution of these galaxies we find that a number are redder than the two BCG's ($u-r\sim3$), thus we additionally remove any galaxy with $u-r>3.5$. We define the 350 galaxies that pass these selection cuts as cluster members.

\begin{figure}
\centering
\includegraphics[width=.5\textwidth]{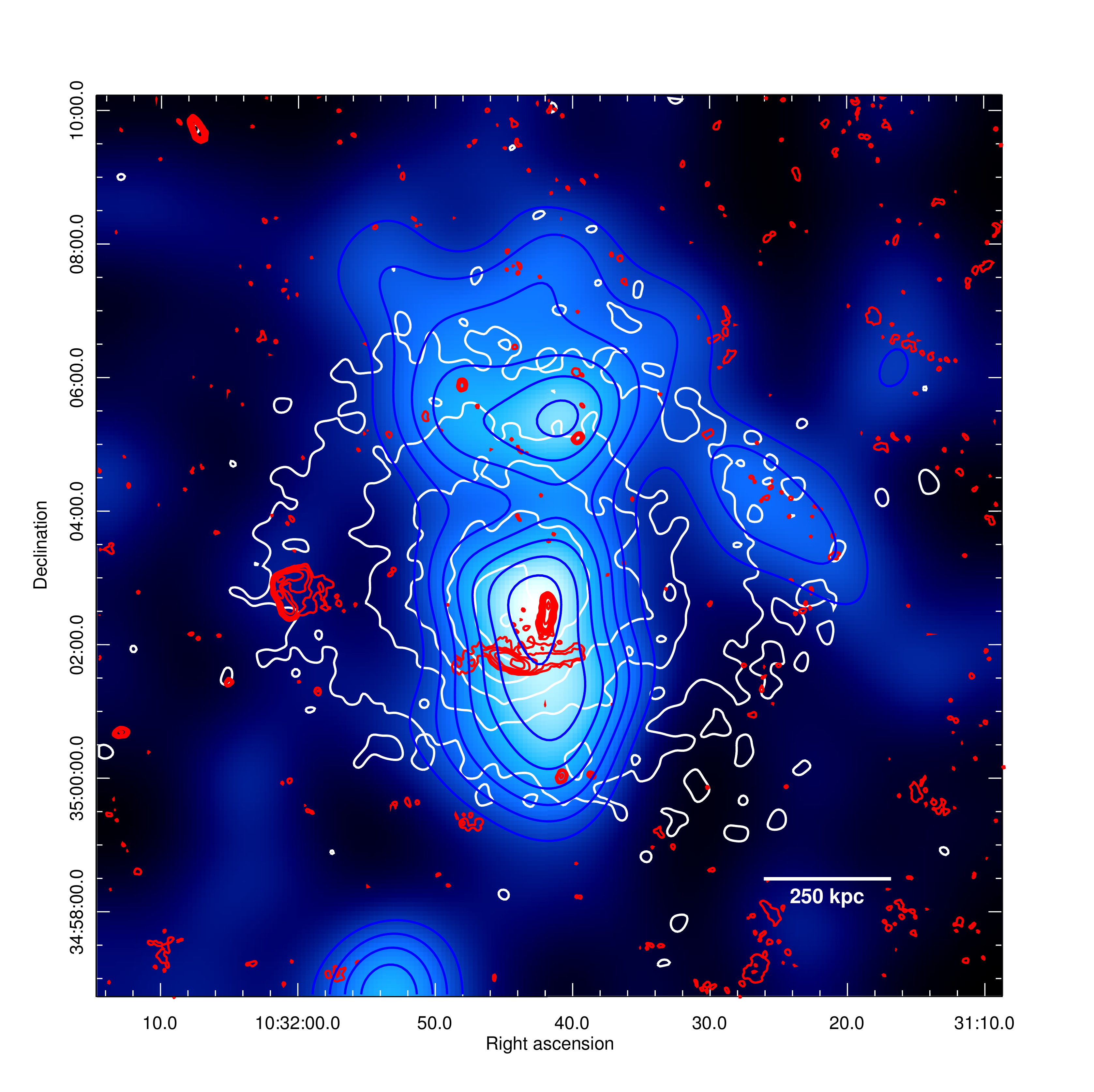}
\caption{\target{} galaxy number density map (blue scale), based on the cluster member selection cuts described in \S\ref{sec:galdist}.
The blue contours begin at 100\,galaxies\,Mpc$^{-2}$ and increase in increments of 25\,galaxies\,Mpc$^{-2}$.
The distribution is largely bimodal showing evidence for a north and south subcluster.
The X-ray gas (white contours; from Fig.~\ref{fig:Chandra}) is peaked near the southern subcluster, while there is no apparent gas peak near the northern subcluster.
The VLA radio contours from Fig.~\ref{fig:radio-vla} are shown in red.}\label{fig:sdss_distrib} 
\end{figure}

We plot the distribution of cluster members, after smoothing with a Gaussian kernel with $\sigma=41\arcsec$, in Fig.~\ref{fig:sdss_distrib}. We find that the distribution of cluster galaxies is largely bimodal with evidence for a north and south subcluster. Both north and south BCGs (BCG north: RA \hms{10}{31}{39.8} -- DEC \dms{+35}{05}{34.4}, BCG south: RA \hms{10}{31}{44.3} -- DEC \dms{+35}{02}{29.0}) are within 50~kpc projected distance form the relative subcluster number density peaks (subcluster north: RA \hms{10}{31}{40.8} -- DEC \dms{+35}{05}{27}, subcluster south: RA \hms{10}{31}{42.7} -- DEC \dms{+35}{02}{41}). The X-ray luminosity peak (RA \hms{10}{31}{44.8} -- DEC \dms{+35}{02}{18}) is within 100~kpc projected distance from the southern subcluster number density peak, but it is very close to the southern BCG. Despite the similar extent and galaxy densities of the north and south subclusters, there is no apparent X-ray peak associated with the northern one.

To examine the significance of the north and south subclusters we created 1000 bootstrap realizations of the galaxy number density map. Using the variance of these bootstrap realizations and assuming Gaussian statistics we find that the southern subcluster has a peak signal-to-noise-ratio (SNR) of 5.6, the northern subcluster has a peak SNR of 4.9, the 150 galaxies\,Mpc$^{-2}$ level has an average SNR $\approx4.0$, and the trough region between the two peaks has an average SNR $\approx4.5$. We also find the same structures in a luminosity weighted density map. Thus we believe \target{} is composed of two subclusters of comparable richness.

\subsubsection{Redshift Analysis}
\label{sec:redshift_analysis}

\begin{figure}
\centering
\includegraphics[width=.5\textwidth]{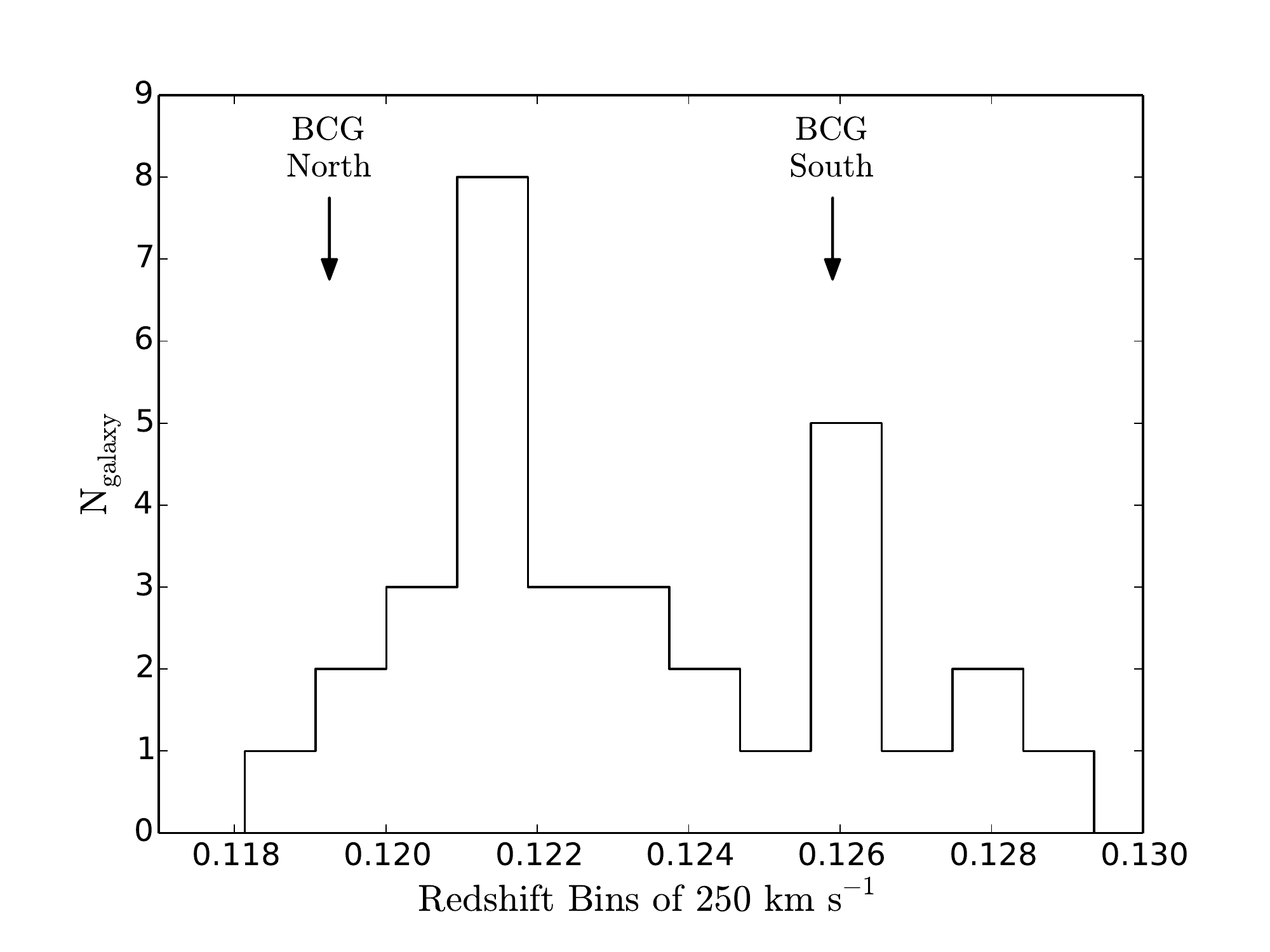}
\caption{Redshift distribution of the 33 spectroscopic galaxies within $\pm3000$km\,s$^{-1}$ of the mean redshift of $z=0.1230\pm0.0005$. The redshift of the north and south BCG's are denoted by the arrows. While there appears to be a bimodal redshift distribution it is unlikely that the observed distribution is an accurate representation of the full cluster redshift distribution given the complex nature of the cluster and limited number of spectroscopic redshifts.}\label{fig:redshifthist} 
\end{figure}

We examined the redshift properties of the 33 spectroscopic cluster galaxies. We show the redshift distribution of these galaxies in Fig.~\ref{fig:redshifthist}.
For these galaxies we estimate a redshift of $z=0.1230\pm0.0005$ and velocity dispersion of $800\pm80$\,km\,s$^{-1}$ using the biweight-statistic and bias-corrected 68\% confidence limits \citep{Beers1990} applied to 100,000 bootstrap samples of the cluster's spectroscopic redshifts.
Given the bimodal galaxy number density (Fig.~\ref{fig:sdss_distrib}) and disturbed redshift distribution (Fig.~\ref{fig:redshifthist}) it is likely that the velocity dispersion is biased.
The redshift distribution, Fig.~\ref{fig:redshifthist}, shows signs of being bimodal, however we note that it is unlikely that the observed distribution is an accurate representation of the full cluster redshift distribution given the complex nature of the cluster and limited number of spectroscopic redshifts.

To investigate whether there is a correlation between the bimodal structure in the projected galaxy number density (Fig.~\ref{fig:sdss_distrib}) and redshift distribution (Fig.~\ref{fig:redshifthist}) we perform both a Dressler-Shectman test \citep[DS-test;][]{Dressler1998} and a simple redshift analysis of the spectroscopic galaxies near each subcluster.
We perform the DS-test using $\lceil\sqrt{33}\rceil$ nearest neighbours and find no evidence of subclustering. 
For the simple redshift analysis of the north and south subclusters we select the spectroscopic galaxies within a radius of 160$\arcsec$ ($\sim$350\,kpc) of each respective subcluster number density peak.
These circular apertures are as large as possible while maintaining mutual exclusivity of the subcluster spectroscopic members.
These selections resulted in 9 galaxies being assigned to each subcluster.
We perform the same redshift analysis as done for the system and find the redshifts of the north and south subclusters to be $0.1240^{+0.0008}_{-0.0009}$ and $0.1224^{+0.0008}_{-0.0008}$, respectively, and the velocity dispersions\footnote{Again, we caution that both the redshift estimates and velocity dispersion estimates of the subclusters are likely to be biased and, as discussed below, this bias is expected to be large.} to be $720^{+190}_{-100}$\,km\,s$^{-1}$ and $690^{+150}_{-90}$\,km\,s$^{-1}$, respectively.
We find that the calculated redshift of each subcluster is located between the peaks of the redshift distribution (Fig.~\ref{fig:redshifthist}) suggesting that the redshift estimate of each subcluster may be biased due to the mixing of two redshift populations.

\begin{figure*}
\centering
\subfloat[Cluster]{\includegraphics[width=.5\textwidth]{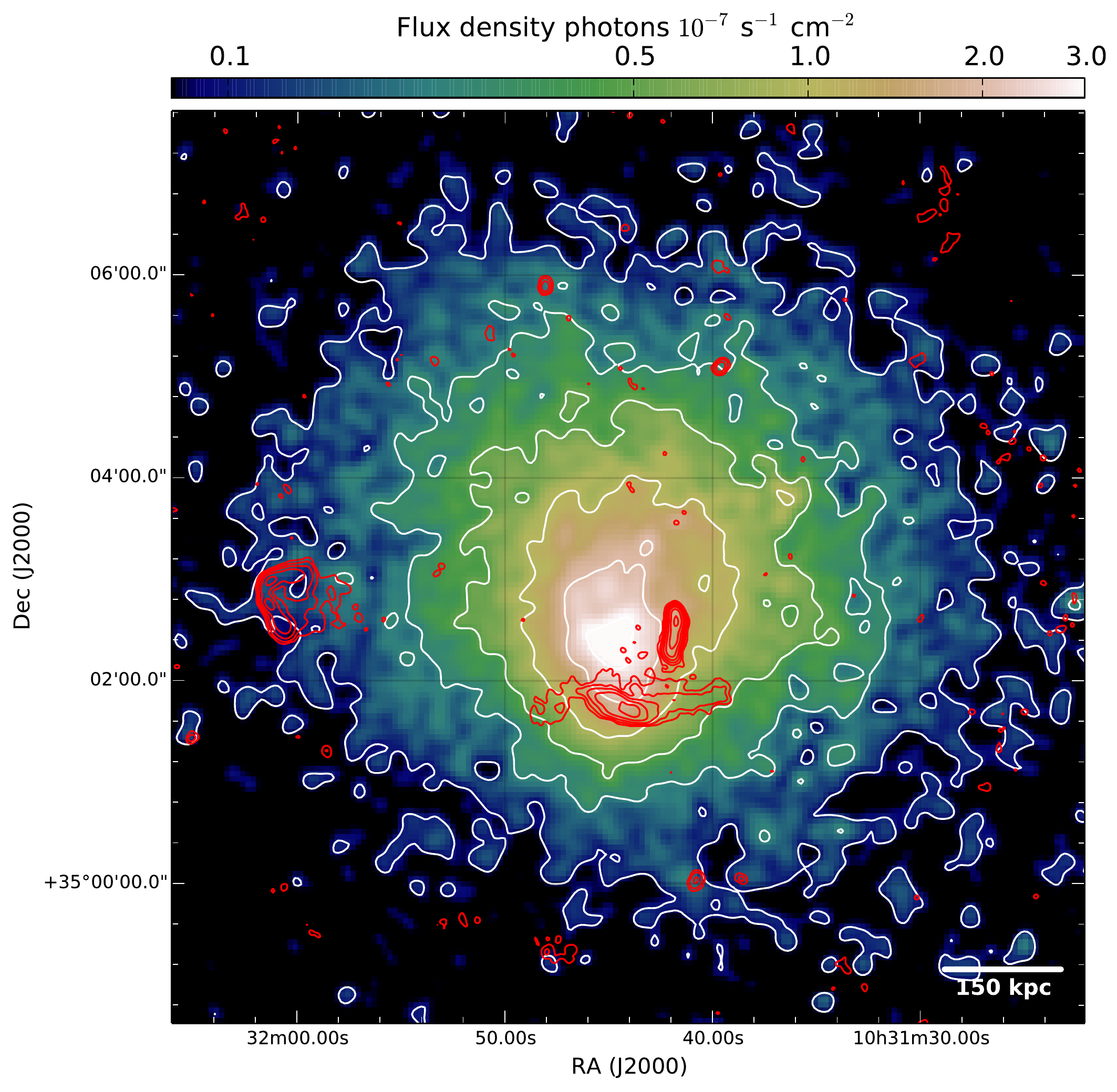}\label{fig:Chandra}}
\subfloat[Zoom-in]{\includegraphics[width=.5\textwidth]{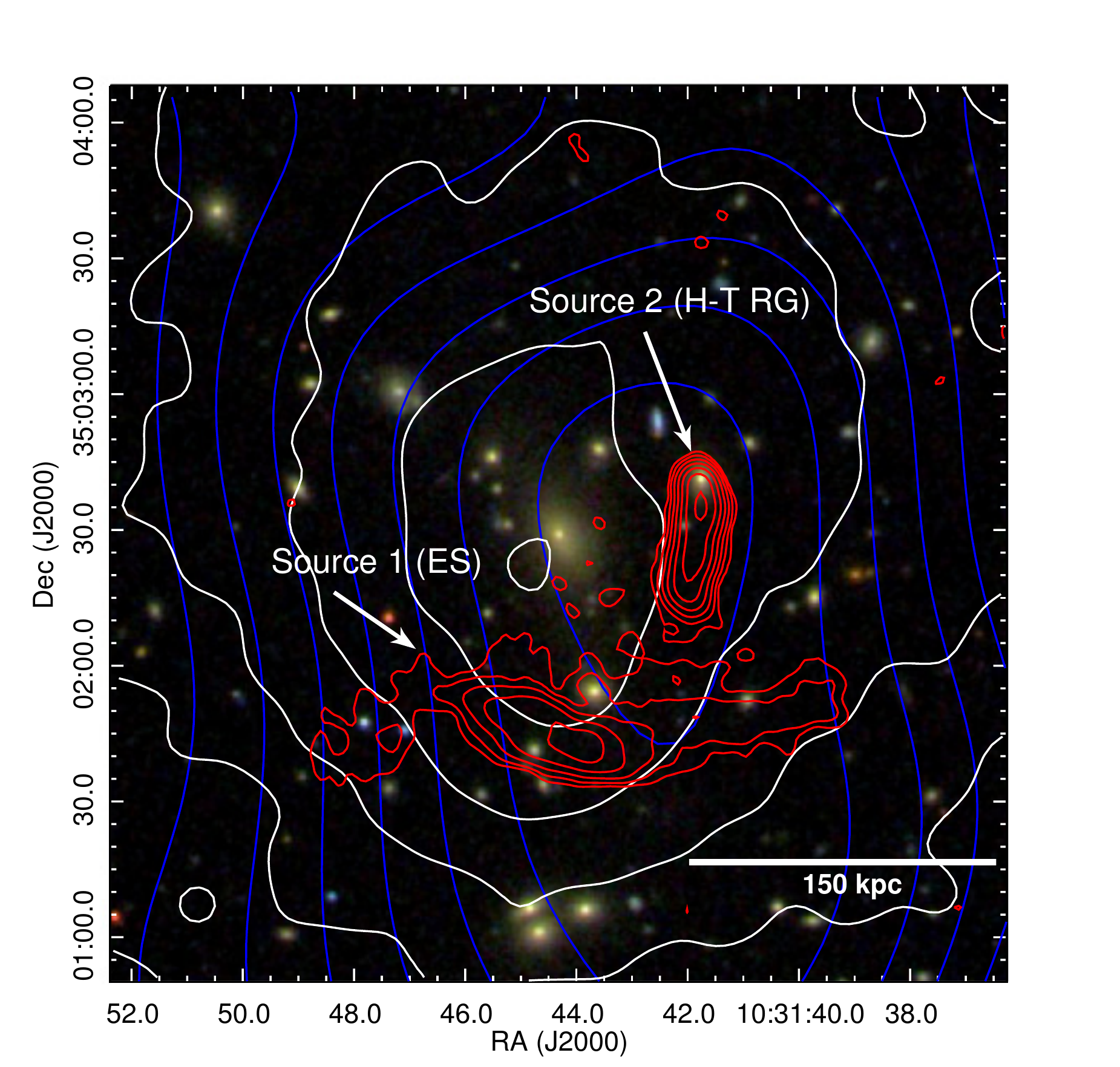}\label{fig:sdss2}}\\
\subfloat[Radio source 3]{\includegraphics[width=.5\textwidth]{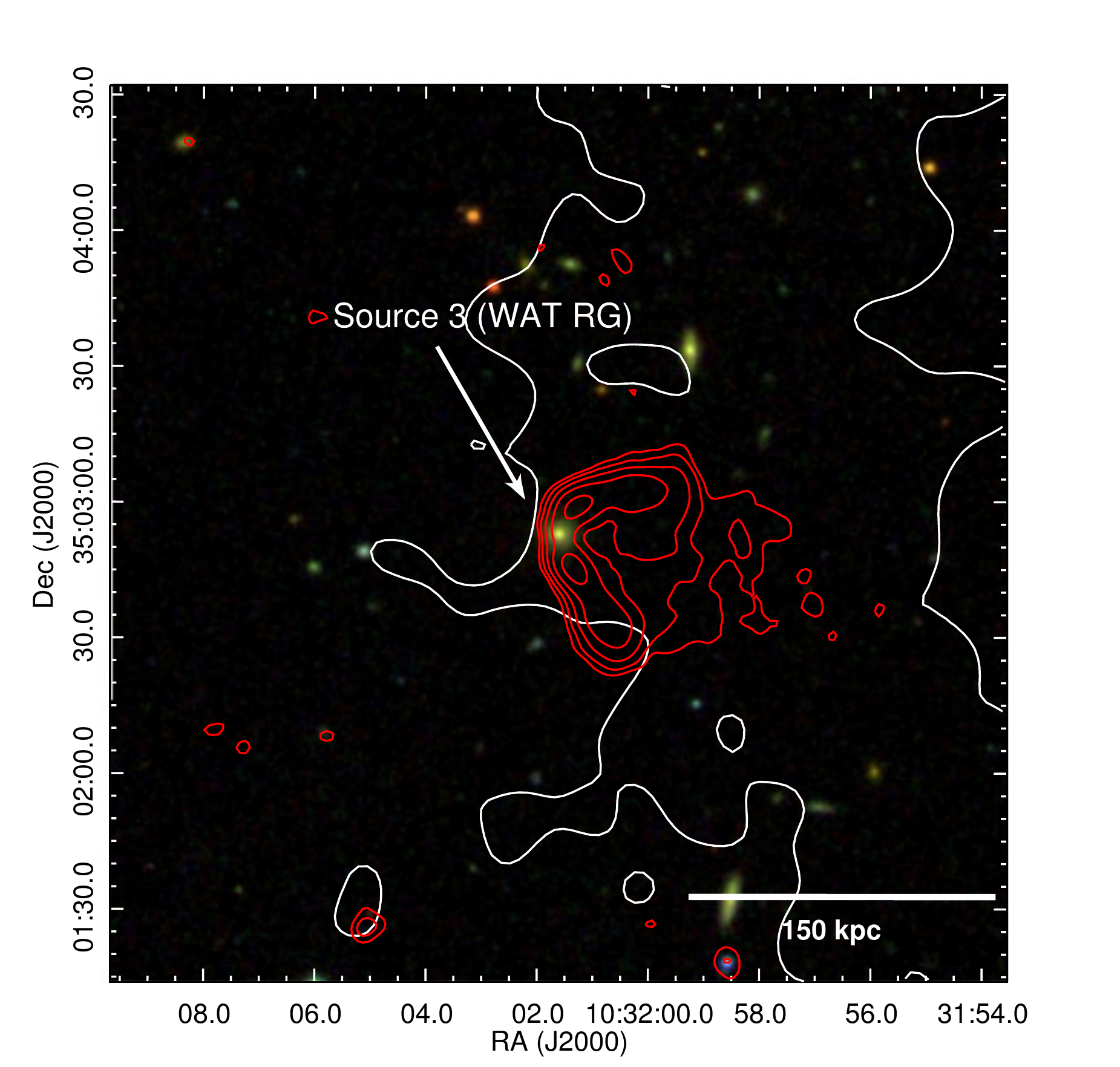}\label{fig:sdss3}}
\subfloat[Radio source 4]{\includegraphics[width=.5\textwidth]{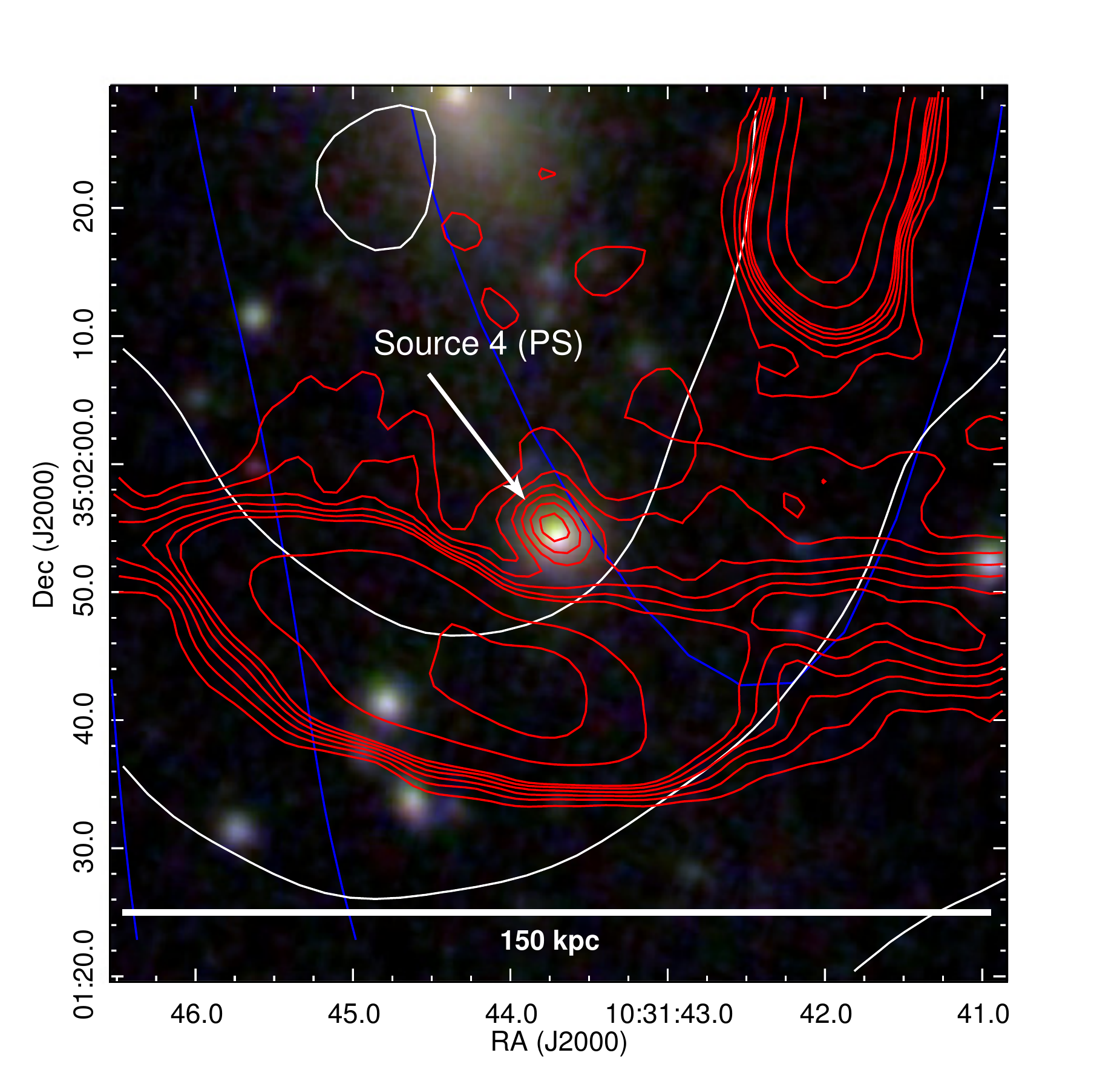}\label{fig:sdss4}}
\caption{Top-left panel: \emph{Chandra} surface brightness map, in the energy band $0.5-4$~keV. The image was binned by 4, exposure- and vignetting-corrected, and smoothed with a two-dimensional Gaussian kernel of $\sigma=2$~pixels (1~$\textrm{pixel} = 1.97$~arcsec). Points sources were subtracted from the image, and each resulting gap was filled with values drawn from a Poisson distribution with a mean equal to that of the pixel values in a small elliptical annulus surrounding the gap. Because this image is highly-processed, it has not been used in the scientific interpretation. Contours are at $\left[0.1,0.25,0.5,1,2,3,8\right] \times 10^{-7}$ s$^{-1}$ cm$^{-2}$. Red contours are from the VLA observation as described in Fig.~\ref{fig:radio}. Other panels: Sloan digital sky survey composite image ($i$, $r$, and $g$ filters). Blue contours trace the projected galaxy density distribution as in Fig.~\ref{fig:sdss_distrib}. White contours trace X-ray emission as shown in the top-left panel. Red contours are from the VLA observation (Fig.~\ref{fig:radio-vla}) at $\left[1, 2, 4, 8, 16, 32, 64\right] \times 3 \sigma$, with $\sigma=47$~\mujybeam. In the last image contours at $\left[1.5,2.5,3.5\right] \times 3 \sigma$ are also plotted. The optical counterpart of the head-tail and the wide angle tail radio galaxies are detected, no optical counterpart for the extended emission (Source 1) is instead detected.}\label{fig:sdss}
\end{figure*}

Comparing the relative line-of-sight velocities\footnote{We use a relativistic velocity convention \citep[see e.g.,][]{Dawson2013}.} of the BCG's and subclusters provides further evidence that there are two overlapping redshift distributions.
We find a large relative velocity difference between the north and south BCG's, $v_\mathrm{north, BCG}-v_\mathrm{south, BCG}=-1776$\,km\,s$^{-1}$, however we find a much smaller relative velocity difference, and of the opposite sign, between the north and south subclusters, $v_\mathrm{north, subcluster}-v_\mathrm{south, subcluster}=500\pm300$\,km\,s$^{-1}$.
Similarly we find large relative velocities between each subcluster and its respective BCG, $v_\mathrm{north, subcluster}-v_\mathrm{north, BCG}=1300\pm200$\,km\,s$^{-1}$ and $v_\mathrm{south, subcluster}-v_\mathrm{south, BCG}=-900\pm200$\,km\,s$^{-1}$.
Even if this is a post merger cluster it seems unlikely that BCG's could receive a large enough dynamical kick to have such large line-of-sight velocities relative to their subclusters while remaining near the subcluster centres in projected space.
A more plausible explanation is that the redshift estimate of each subcluster is biased significantly due to mixing of two subcluster populations with large relative velocities.
Without a more extensive redshift survey we believe that the BCG's provide the least biased estimate of the two subcluster redshifts and relative velocities.
It is worth noting that the northern BCG has a redshift offset from one of the peaks of the observed redshift distributions (Fig.~\ref{fig:redshifthist}), but this may be due to the observed distribution inaccurately representing the full cluster redshift distribution.

\subsection{Radio properties}
\subsubsection{Unclassified extended radio emission}
\label{sec:extended_emission}

Extended radio emission is present 35\arcsec{} below the cluster centre (defined as the peak of the X-ray emission), which corresponds to $\sim 76$ kpc (see Fig.~\ref{fig:radio}). We named this source \textit{Source 1}. The emission extends 281 kpc in length and 39 kpc in width and it does not have any clear optical counterpart (see Fig.~\ref{fig:sdss2}). A \textit{plume} extends from the east side of the source towards south-east. The southern rim of \textit{Source 1} terminates quite sharply while the northern one shows an irregular border with faint extensions elongating towards the cluster's centre. Using \emph{WSRT} and \emph{VLA} data, we investigated the presence of polarized emission from \emph{Source 1}, and found the radiation $<1\%$ polarized in the brightest regions.

The flux densities for \textit{Source 1} extracted from the WSRT and the VLA maps are in agreement in the overlapping frequency range, which means that the VLA observation was able to pick up all the flux although the source was extended and the array quite sparse (B-configuration). The two observations gave a flux of $\sim 48\pm0.5$~\mjybeam{} at 1.4 GHz, which corresponds to a power of $P_{1400} = 1.85 \times 10^{24}$ W Hz$^{-1}$. The properties of \textit{Source 1} are summarized in Table~\ref{tab:ES_proprieties} while flux densities at different frequencies are in Table~\ref{tab:spidx}.

\begin{table}
\centering
\begin{threeparttable}
\begin{tabular}{ccccc}
\hline
Length & Width & Distance from  & Flux    & Power \\
       &       & cluster centre & Density & \\
(kpc)  & (kpc) & (kpc)          & (mJy)   & (W Hz$^{-1}$) \bigstrut[b] \\
\hline
281    & 39    & 76             & $48\pm0.5$ & $1.85 \times 10^{24}$\bigstrut[t]\bigstrut[b]\\
\hline
\end{tabular}
\end{threeparttable}
\caption{Details of the extended source (\textit{Source 1}). Flux density and power are estimated at 1.425 GHz.}\label{tab:ES_proprieties}
\end{table}

\begin{figure}
\centering
\includegraphics[width=.5\textwidth]{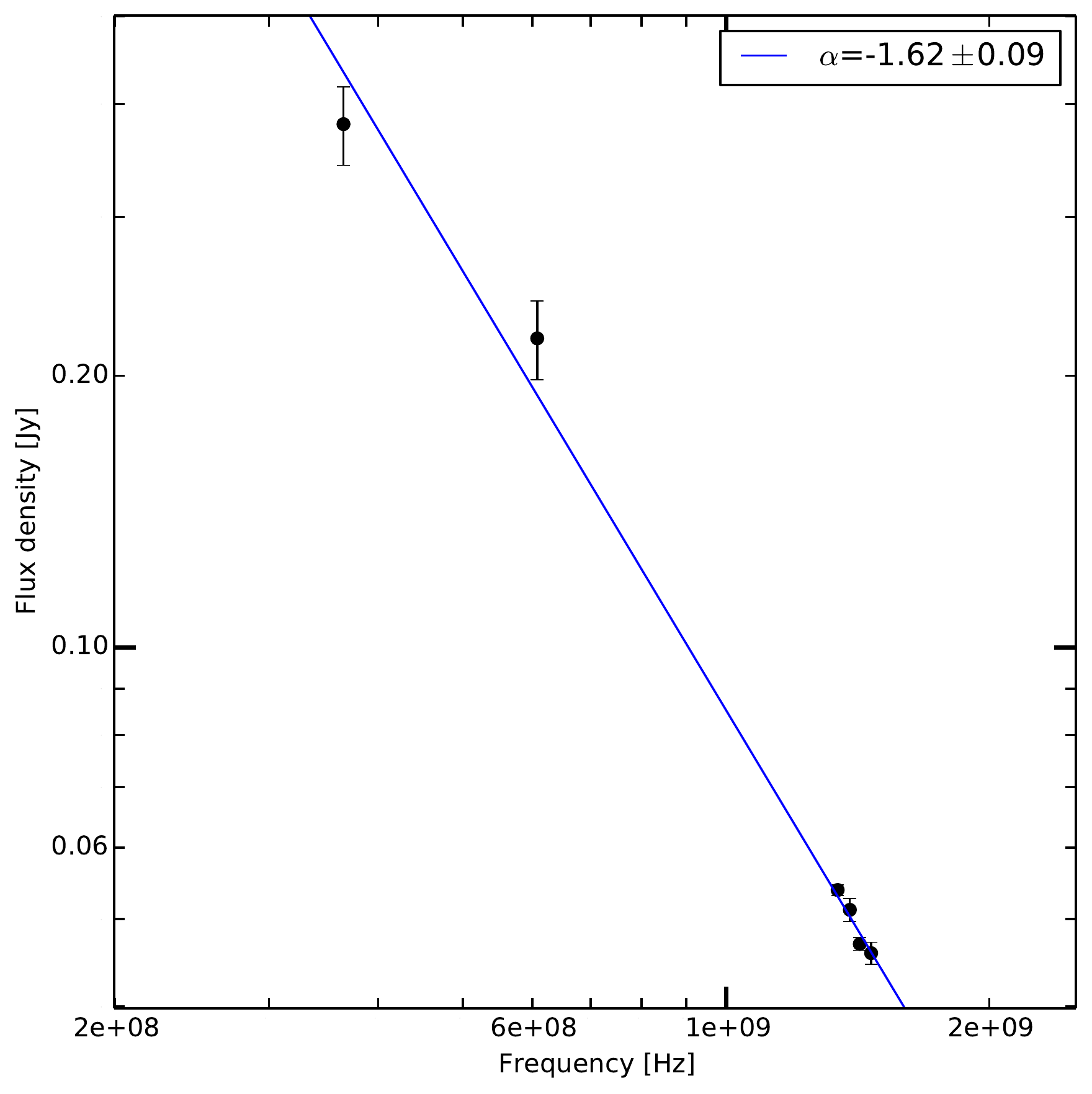}
\caption{Radio spectrum for the extended source (\textit{Source 1}) in \target{}. Data points are listed in Table~\ref{tab:spidx}. Blue line is a linear fit. The errors of the two lowest frequency data-points are unknown. In these two cases, for fitting purposes, we assumed conservative error equal to 10\% of the flux density values.}\label{fig:spidx} 
\end{figure}

This extended radio source was already noticed by \cite{Roland1985} while searching for steep radio sources in galaxy clusters. They found the spectral index of the source to be rather steep ($\alpha < -1.3$, $F_\nu \propto \nu^{\alpha}$), but the authors could not clearly classify it because of a missing optical counterpart. Using data from the literature (see Table~\ref{tab:spidx}) we extracted a global spectral index value for \textit{Source 1} of $\alpha = -1.62\pm0.09$ (see Fig.~\ref{fig:spidx}). Data points were taken from different telescopes covering different $uv$-ranges, which may bias some measurements. However, the spectrum shows a hint of curvature becoming steeper at higher-frequencies, as expected by standard synchrotron ageing mechanisms.

\begin{table}
\centering
\begin{threeparttable}
\begin{tabular}{lccl}
\hline
Telescope & Frequency  & Flux    & Reference \\
          & (MHz)      & (mJy)   & \bigstrut[b] \\
\hline
VLA       & 1465 & $45.8\pm1.3$ & this work \bigstrut[t]\\
WSRT      & 1422 & $46.9\pm7.6$ & this work \\
VLA       & 1385 & $51.2\pm1.5$ & this work \\
WSRT      & 1341 & $53.9\pm7.3$ & this work \\
WSRT      & 608  & 220 & \cite{Roland1985}\\
Texas interferometer & 365  & 380 & \cite{Roland1985}\bigstrut[b]\\
\hline
\end{tabular}
\end{threeparttable}
\caption{Flux densities of the extended sources (\textit{Source 1}).}\label{tab:spidx}
\end{table}

\begin{figure}
\centering
\includegraphics[width=.5\textwidth]{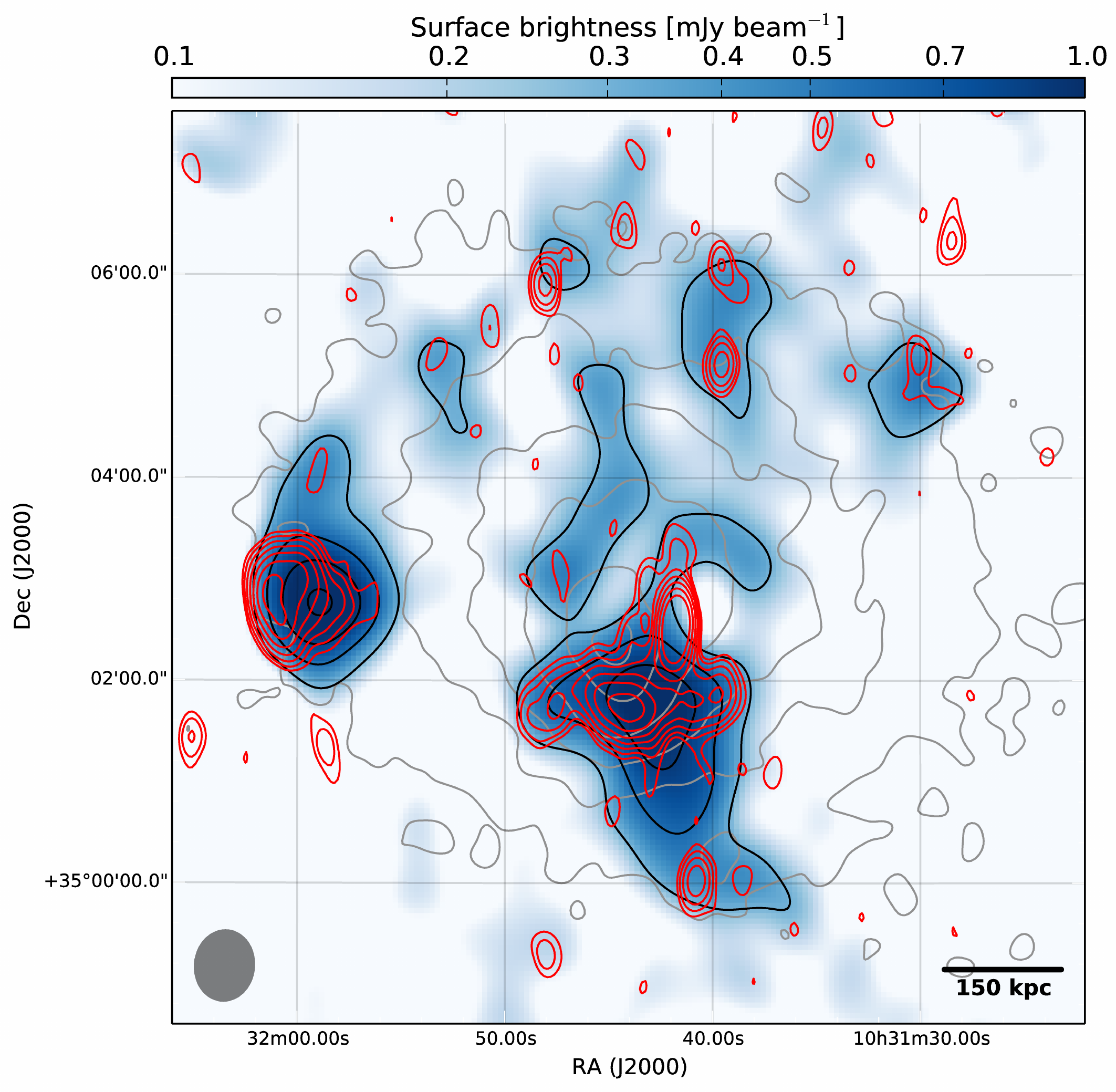}
\caption{WSRT low-resolution map of the \target{} galaxy cluster. The high-resolution image (shown in red contours, Fig.~\ref{fig:radio-wsrt}) has been $uv$-subtracted and data were tapered to the resolution of \beam{41}{34} to enhance the residual extended emission. Contours are at $\left[1,2,3,4\right] \times 3 \sigma$ with $\sigma = 100$~\mujybeam. Gray contours trace the X-ray emission as in Fig.~\ref{fig:Chandra}.}\label{fig:wsrt-ext} 
\end{figure}

To study the presence of a radio halo we examined the low-resolution WSRT map after subtracting the emission of the high-resolution map. The result, displayed in Fig.~\ref{fig:wsrt-ext}, shows an excess of emission throughout the cluster which is close to the noise level. This can be related to the presence of a radio halo, which would not be surprising in a system which underwent a recent merger. However, the emission is too faint to derive firm conclusions. 

\subsubsection{Other radio sources}
\label{sec:other_sources}

The radio environment of \target{} is quite complex. A head-tail (H-T) radio galaxy (RG), oriented north-south, is present close to the cluster centre. It has a clear optical counterpart 2MASX J10314175+3502411 ($z=0.120$) which is probably part of the northern sub-cluster of \target{}. The source has been labelled \textit{Source 2} (see Fig.~\ref{fig:sdss2}).

\begin{figure}
\centering
\includegraphics[width=.5\textwidth]{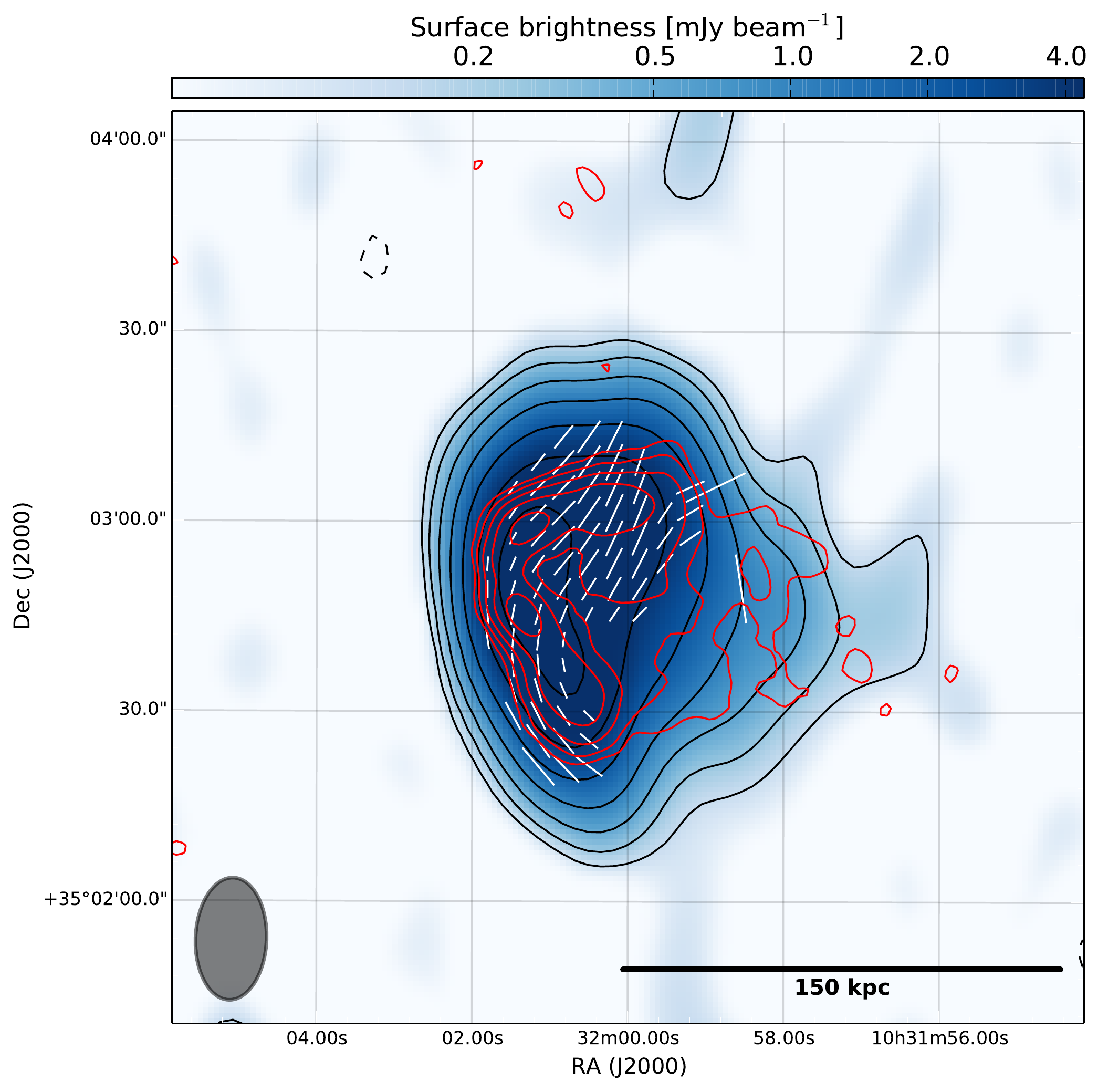}
\caption{WSRT 1.4 GHz map of the WAT source NVSS J103200+350249. Black contours are at $\left[1, 2, 4, 8, 16, 32, 64\right] \times 3 \sigma$, with $\sigma=44$~\mujybeam, red contours are from the VLA 1.4 GHz observation at $\left[1, 2, 4, 8, 16, 32\right] \times 3 \sigma$, with $\sigma=47$~\mujybeam. E-Vectors are displayed with length proportional to the fractional polarization which ranges from 2 to 15\%. A correction for Galactic Faraday rotation has been applied following \citet{Oppermann2012}.}\label{fig:wsrt-wat} 
\end{figure}

A wide angle tail (WAT) RG is present on the east side of the cluster and it is labelled \textit{Source 3} (see Fig.~\ref{fig:sdss3}). An optical counterpart for this object has been identified as 2MASX J10320162+3502530 ($z=0.121$) which is also part of the northern sub-cluster of \target{}. The emission from the WAT RG is polarized up to 15\%. Fig.~\ref{fig:wsrt-wat} shows the orientation of the E-vectors (perpendicular to the magnetic fields) for this source. Using the map provided by \cite{Oppermann2012} we estimated the Faraday depth in the direction of \textit{Source 3} to be $\sim9.3$ rad m$^{-2}$. The polarization angle has therefore been corrected for Galactic Faraday rotation by subtracting an angle of 25\deg. 

A fourth radio source is interesting for further discussion. It is a faint ($S_{1400} = 0.5$~mJy) point source (PS) with an optical counterpart (SDSS J103143.69+350154.5, $z=0.125$) located in projection 29 kpc North of the brightest peak of \textit{Source 1}. Given its location and redshift the source is probably part of southern sub-cluster of \target{} and it is labelled \textit{Source 4} (see Fig.~\ref{fig:sdss4}). A list of the proprieties of the sources discussed here is given in Table~\ref{tab:sources}.

\begin{table*}
\centering
\begin{threeparttable}
\begin{tabular}{lccccl}
Source name & RA\tnote{a}\ \ (J2000) & Dec\tnote{a}\ \ (J2000) & $S_{1.425~{\rm GHz}}$ (mJy) & z\tnote{b} & Note\bigstrut[b]\\
\hline
FIRST J103143.9+350141 & \hms{10}{31}{43} & \dms{+35}{01}{42} & $48.0\pm0.5$ & -- & Extended source (Source 1) \bigstrut[t]\\
FIRST J103141.7+350231 & \hms{10}{31}{41.77} & \dms{+35}{02}{41.42} & $46.7\pm0.2$ & 0.120 & Head-tail RG (Source2 )\\
NVSS J103200+350249 & \hms{10}{32}{01.61} & \dms{+35}{02}{52.94} & $46.8\pm0.5$ & 0.121 & WAT RG (Source 3) \\
SDSS J103143.69+350154.5 & \hms{10}{31}{43.69} & \dms{+35}{01}{54.50} & $0.5\pm0.05$ & 0.125 & Point source (Source 4)\\
\end{tabular}
\begin{tablenotes}
\item[a] coordinates are from the SDSS optical counterpart. In the case of the extended source, they are at the peak of the radio emission.
\item[b] spectroscopic redshift from the SDSS optical counterpart.
\end{tablenotes}
\end{threeparttable}
\caption{Coordinates and fluxes of the radio sources.}\label{tab:sources}
\end{table*}

\subsection{X-ray properties}
\subsubsection{X-ray global parameters}
\label{sec:globalXray}

To determine the global cluster parameters, we extracted cluster spectra from a circular region of radius $5.3^\prime$ ($\approx 690$~kpc) centred at RA = \hms{10}{31}{44.4} and DEC = \dms{+35}{03}{12.0}. This region corresponds approximately to the cluster's $R_{500}$, and encloses the main part of the visible X-ray emission from the ICM. From the total spectra, we subtracted instrumental background spectra extracted from the same regions. The sky background parameters were kept fixed, in turn, to each of the two best-fitting models summarized in Table \ref{tab:skybkg}. Emission from the ICM was modelled with a single-temperature APEC model with free temperature, metallicity, and normalization, and with a fixed redshift of $0.1259$. The spectra from ObsIDs 15084 and 15614 were fitted simultaneously, under the assumption that they are described by the same model parameters. The fits were done in the $0.5-7$~keV energy band.

\begin{table*}
\centering
\begin{threeparttable}
\begin{tabular}{cccccc}
Sky background & $\alpha_2$ &  $\alpha_1$ & $r_{\rm d}$ & $C$  & FCN\tnote{a} \\
(photons~s$^{-1}$~cm$^{-2}$~arcmin$^{-2}$) & & & (arcmin) & &  \bigstrut[b]\\
\hline
$5.92\times 10^{-6}$ & $0.93_{-0.18}^{+0.17}$ & $1.63\pm 0.12$ & $2.25_{-0.03}^{+0.05}$ & $1.42_{-0.16}^{+0.18}$ & $-1548.07$ \bigstrut[t]\\
$6.60\times 10^{-6}$ & $0.93_{-0.18}^{+0.17}$ & $1.67\pm 0.13$ & $2.25_{-0.03}^{+0.05}$ & $1.41_{-0.16}^{+0.18}$ & $-1547.99$ \\
$7.30\times 10^{-6}$ & $0.94_{-0.18}^{+0.17}$ & $1.72_{-0.13}^{+0.14}$ & $2.25_{-0.03}^{+0.05}$ & $1.40_{-0.16}^{+0.18}$ & $-1547.88$ \\
\end{tabular}
\begin{tablenotes}
\item[a] \url{http://wwwasdoc.web.cern.ch/wwwasdoc/minuit/node14.html}
\end{tablenotes}
\end{threeparttable}
\caption{Best-fitting broken power-law density models to the surface brightness profile extracted from the sector shown in Fig.~\ref{fig:sxsector}. The sky background value was kept fixed in the fit. We ran the fit for the best-fitting sky background level of $6.60\times 10^{-6}$~photons~s$^{-1}$~cm$^{-2}$~arcmin$^{-2}$, and for sky background levels at the boundaries of the $90\%$ confidence interval.}\label{tab:bknpow}
\end{table*}

\begin{figure}
\centering
\includegraphics[width=\columnwidth]{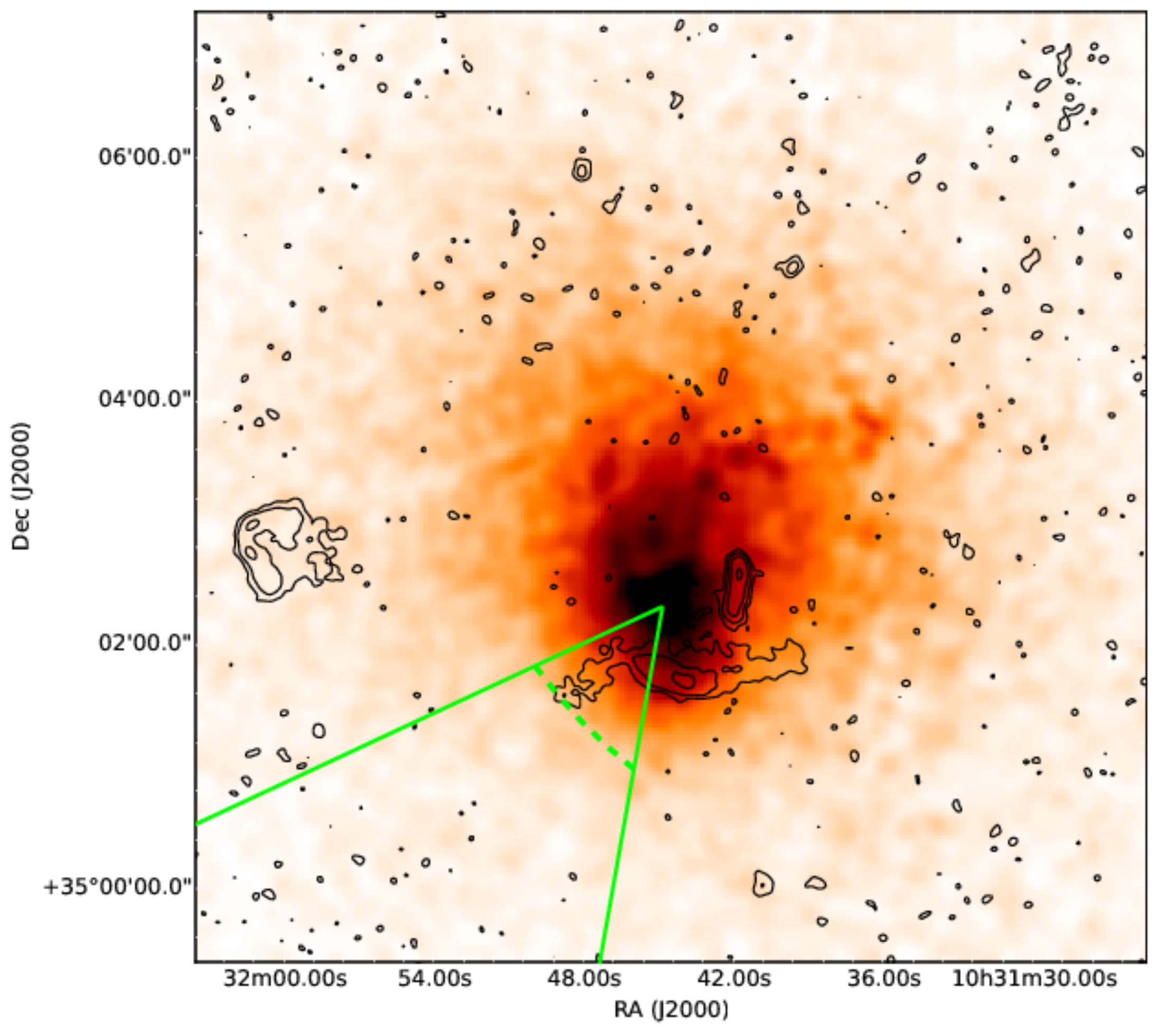}
\caption{\emph{Chandra} surface brightness map in the energy band $0.5-3$~keV. The image was binned by 4, exposure-corrected, vignetting-corrected, and smoothed with a two-dimensional Gaussian kernel of $\sigma=2$~pixels (1~pixel = 1.97~arcsec). Point sources were subtracted from the image, and each resulting gap was filled with values drawn from a Poisson distribution with a mean equal to that of the pixel values in a small elliptical annulus surrounding the gap. This is shown for clarity, but because it is highly processed it was not used in the analysis. The overlaid sector shows the region in which a density discontinuity was detected by eye. The discontinuity was modelled with a broken power-law density model, as described in the text and shown in Fig.~\ref{fig:sxfits}. Radio contours based on the 1.4~GHz VLA radio image are drawn at $\left[1,4,16,64\right] \times 100$~\mujybeam.}
\label{fig:sxsector} 
\end{figure}
For the best-fitting sky background model without a hot foreground component, we found an average cluster temperature of $6.34\pm 0.16$~keV and a metallicity of $0.26\pm 0.04$ solar. For the best-fitting sky background model that included a HF component, we found very similar ICM parameters: an average cluster temperaure of $6.39\pm 0.16$ keV, and a metallicity of $0.25\pm 0.04$ solar. The cluster has a $0.5-2$~keV flux of $3.86_{-0.03}^{+0.02} x 10^{-12}$ erg~cm$^{-2}$~s$^{-1}$. The calculated flux is essentially the same for both sky background models, and it implies that Abell~1033 has a $0.5-2$~keV luminosity of $(1.48\pm0.01)\times 10^{44}$~erg~s$^{-1}$. Using the $M_{500} - L$ relation found by \cite{Pratt2009} we can infer a cluster mass of $(3.5\pm0.2) \times 10^{14}$~\Msun.

\subsubsection{Density discontinuities in the ICM}

The main aim of the \emph{Chandra} analysis was to identify density and temperature discontinuities in the ICM, in order to explore their connection with radio sources. The only clear density discontinuity that we were able to detect by eye is located SE of the cluster, in the elliptical sector shown in Fig.~\ref{fig:sxsector}. We extracted a surface brightness profile from this elliptical sector, using images and exposure maps in the energy band $0.5-3$~keV. The instrumental background surface brightness profile extracted from the same sector and in the same energy band was subtracted from the total surface brightness profile, and binned to a minimum of 1 count/bin. We modelled the sky background by fitting a constant to the outer regions ($13-18$ arcmin) of the profile. Once the sky background level was determined, we fitted a broken power-law density model to the surface brightness profile (in the radius range $1-5$~arcmin), keeping the sky background level fixed and assuming isothermal plasma. The broken power-law density model is defined as:
\begin{eqnarray}
	n_2(r) & = & C \, n_0 \, \left(\frac{r}{r_{\rm d}}\right)^{\alpha_2}, \: \: \textrm{for} \; r\le r_{\rm d} \nonumber \\
	n_1(r) & = & n_0 \, \left(\frac{r}{r_{\rm d}}\right)^{\alpha_1}, \: \: \textrm{for} \; r> r_{\rm d}
\end{eqnarray}
where $n$ is the electron number density, $C$ is the density compression, $r$ is the distance from the centre, $r_{\rm d}$ is the distance of the density discontinuity from the centre, $\alpha$ is the power-law index, and subscripts 1 and 2 refer to the outer side of the density discontinuity and inner side of the density discontinuity, respectively. The density profile is integrated along the line of sight under the assumption that the cluster's 3D geometry is a prolate spheroid. If the cluster would instead be longer along the line of sight, then the density jump we detect would be stronger.

We also varied the sky background level within its $90\%$ confidence limits, and repeated the fitting. All fits were done using \textsc{proffit}~v1.2 \citep{Eckert2011} and Cash statistics \citep{Cash1979}.

\begin{figure}[ht!]
\centering
\includegraphics[width=0.5\textwidth]{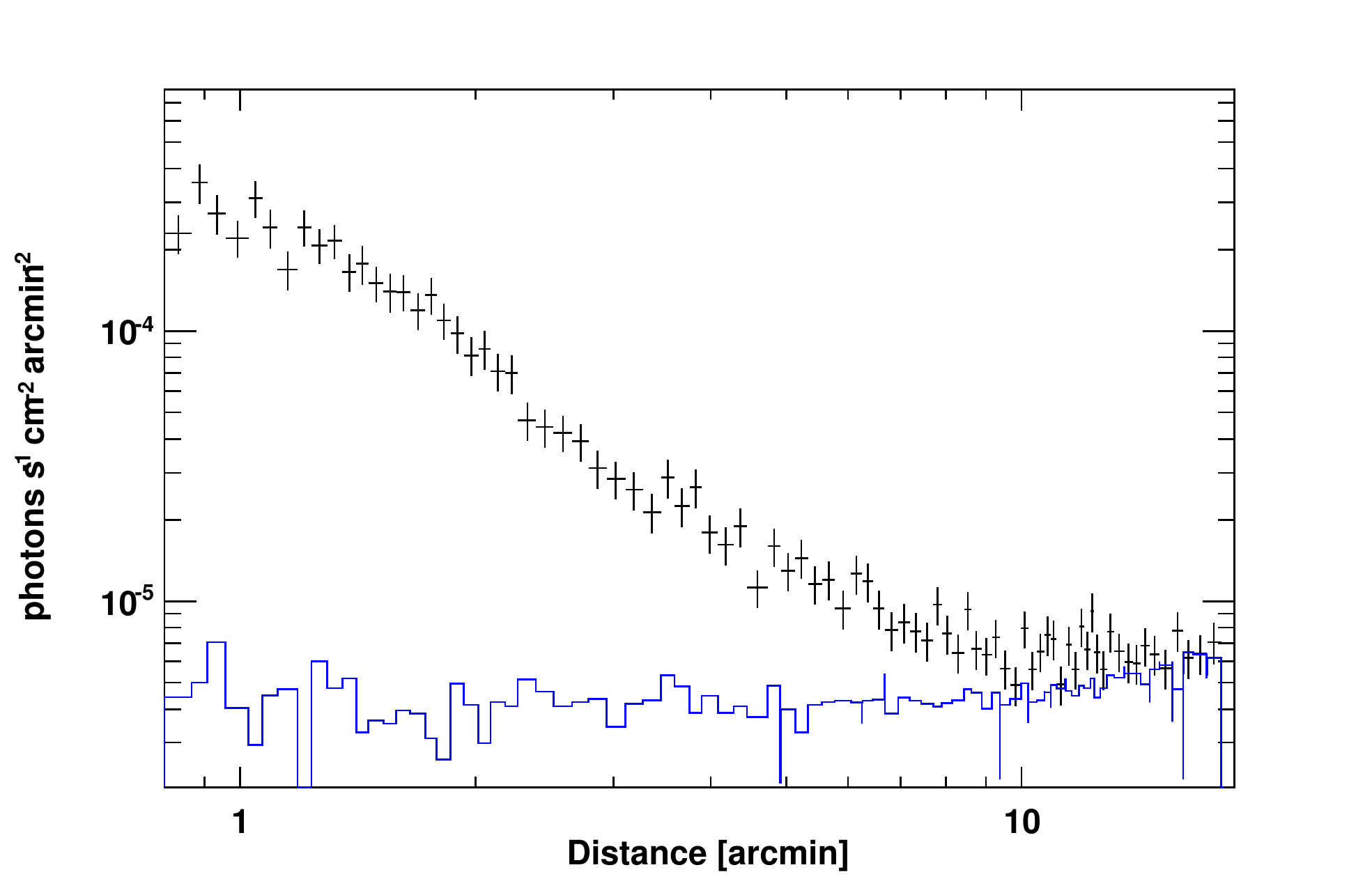}\\
\includegraphics[width=0.5\textwidth]{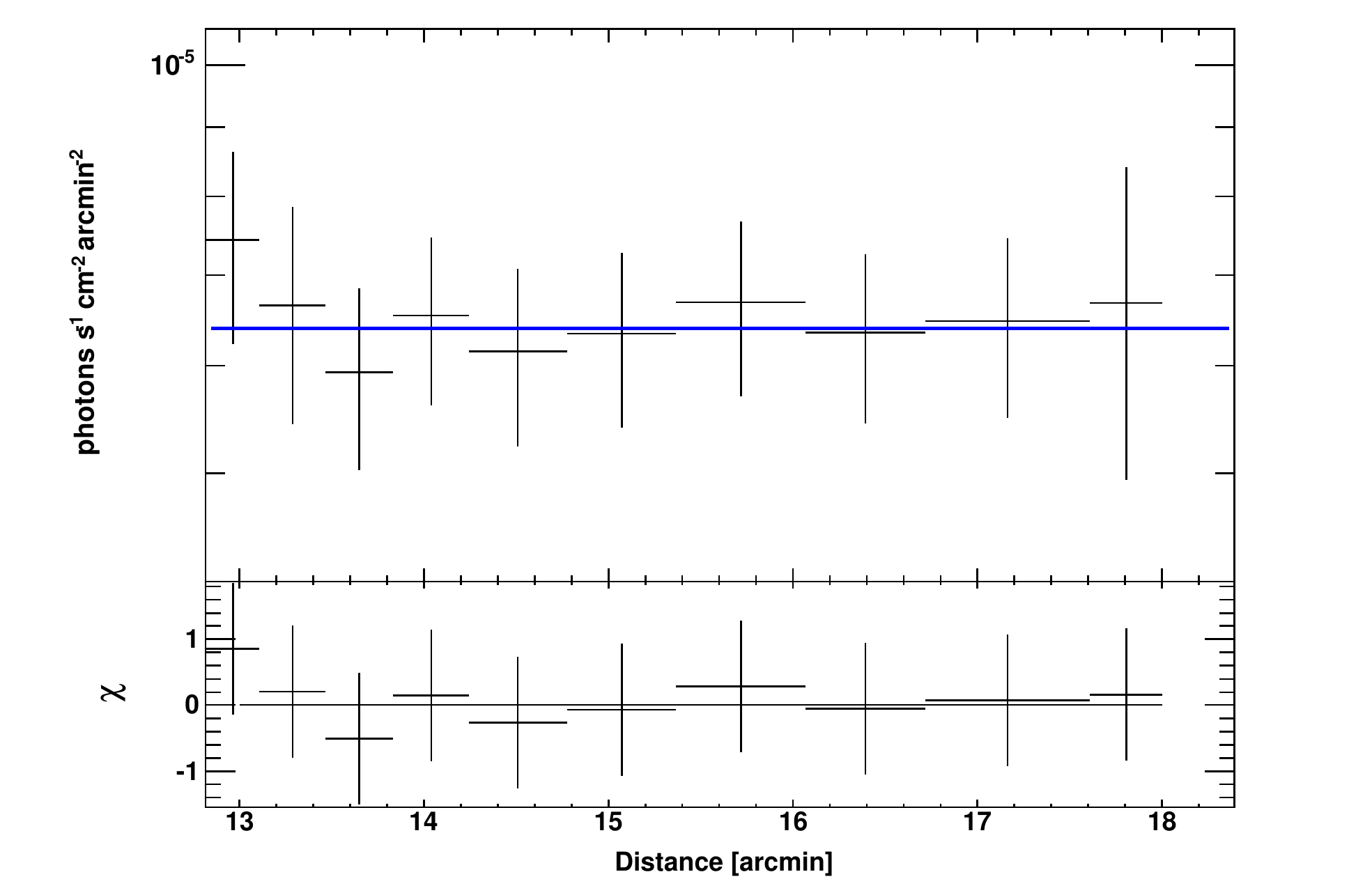}\\
\includegraphics[width=0.5\textwidth]{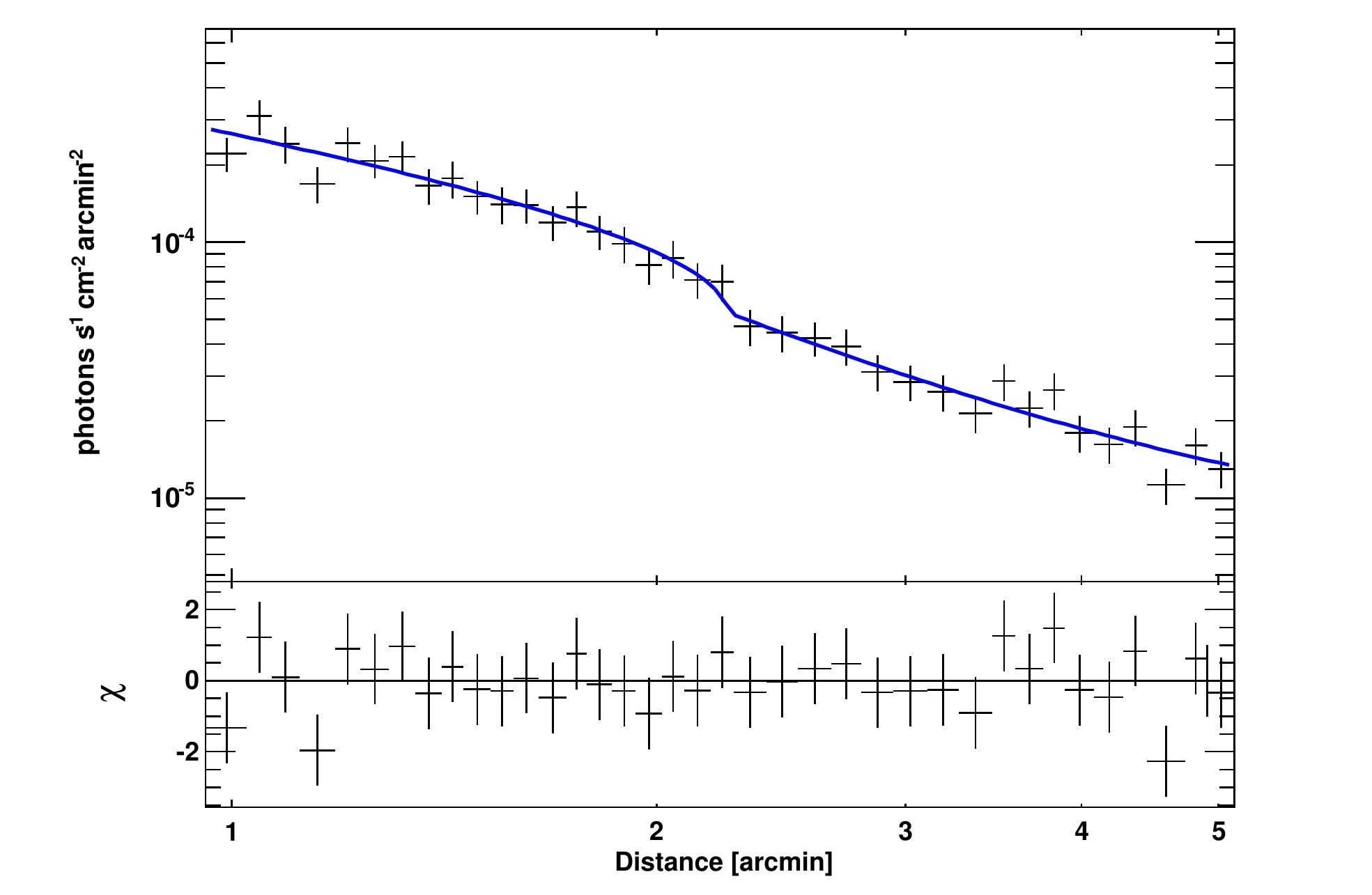}
\caption{\emph{Left:} surface brightness profile extracted from the sector shown in Fig.~\ref{fig:sxsector}. For clarity, the profile was binned to a uniform signal-to-noise ratio of 6. The blue line shows the instrumental background profile; in the plot, this is subtracted from the total profile. \emph{Middle}: constant model fit to the sky background. For clarity, the profile was binned to approximately 30 counts/bin. The fitted model is shown in blue. The bottom panel shows the fit residuals. \emph{Right}: broken power-law density model fit to the radial range $1-5$~arcmin. For clarity, the profile was binned to a uniform signal-to-noise ratio of 6. The fitted model is shown in blue. The bottom panel shows the fit residuals.}
\label{fig:sxfits} 
\end{figure}

For the sector shown in Fig.~\ref{fig:sxsector}, the sky background level in the energy band $0.5-3$~keV is $6.60_{-0.41}^{+0.42}\times 10^{-6}$~photon~s$^{-1}$~cm$^{-2}$~arcmin$^{-2}$ (errors are quoted at the $1\sigma$ level; the $90\%$ confidence range is $[5.92-7.30]\times 10^{-6}$~photon~s$^{-1}$~cm$^{-2}$~arcmin$^{-2}$). The best-fitting broken power-law models for different sky background values are summarized in Table~\ref{tab:bknpow}. The background and ICM fits to the surface brightness profiles are shown in Fig.~\ref{fig:sxfits}. The best-fitting broken power-law model parameters are essentially unchanged as the sky background level is varied. The best-fitting broken power-law model derived for the best-fitting sky background level has a density compression $C=1.41_{-0.16}^{+0.18}$, which implies a very weak Mach number of $1.28_{-0.11}^{+0.12}$ if the discontinuity is a shock front. The position of the density discontinuity is shown in Fig.~\ref{fig:sxsector}. The broken power-law fit was compared to a kink model fit (i.e. a broken power-law model with a fixed density compression $C=1$) and to a simple power-law fit. Based on the likelihood-ratio test, a kink in the profile is confirmed at $4.2\sigma$ (confidence level $99.997\%$), while a density compression $C>1$ is confirmed at $1.82\sigma$ (confidence level $93.197\%$).

We searched for other density discontinuities in the ICM using narrow ($\sim 20-30$ degree) circular sectors that approximately follow the X-ray isophotes. However, no other discontinuities were detected with a confidence level $\ge 90\%$.

\subsubsection{Spectral analysis near the density discontinuity}

To identify the nature of the density discontinuity detected $\approx$~\arcmind{2}{25} SE of the cluster centre, we extracted spectra in two partial elliptical annuli on both sides of the discontinuity. The two annuli have widths of approximately 0.7 arcmin (inner region) and 1.1~arcmin (outer region). A negative temperature jump would indicate that the discontinuity corresponds to a shock front. Instead, if the temperature increases significantly on the outside of the density discontinuity, then the discontinuity would correspond to a cold front.

The two spectra were binned to a minimum of 25 counts/bin. To fit the spectra, we subtracted the instrumental background spectra and fixed the sky background model to the best-fitting models summarized in Table~\ref{tab:skybkg}. The ICM was modelled with a single-temperature APEC model with a fixed metallicity of $0.25$ solar and a fixed redshift of 0.1259. The hydrogen column density was fixed to the Galactic value of $1.64\times 10^{20}$~atoms~cm$^{-2}$. The spectra from ObsID 15084 and 15614 were fitted simultaneously, under the assumption that they are described by the same physical model. The results of the fits are summarized in Table \ref{tab:tjump}.

\begin{table*}
\begin{center}
\begin{tabular}{lcccccc}
Sky background	& $T_2$	& $\mathcal{N}_2$  & $\chi^2/{\rm d.o.f.}$	&  $T_1$	& $\mathcal{N}_1$ & $\chi^2/{\rm d.o.f.}$	\\
model		& (keV) & (cm$^{-5}$~arcmin$^{-2}$) &  & (keV) & (cm$^{-5}$~arcmin$^{-2}$) & \\
\hline
with HF	&  $6.15_{-0.88}^{+1.14}$ & $(3.95\pm 0.15)\times 10^{-4}$ & $0.99$ & $5.57_{-1.11}^{+1.69}$ & $6.41_{-0.36}^{+0.39}\times 10^{-5}$ & $0.66$ \bigstrut[t]\\
without HF&  $6.13_{-0.88}^{+1.13}$ & $(3.95\pm 0.15)\times 10^{-4}$ & $0.99$ &  $5.42_{-1.04}^{+1.64}$ & $6.41_{-0.36}^{+0.39}\times 10^{-5}$ & $0.67$ \bigstrut[b]\\
\end{tabular}
\caption{Best-fitting APEC models to the spectra extracted from both sides of the SE density discontinuity. The fits were run with the sky background model fixed, in turn, to each of the two best-fitting models summarized in Table~\ref{tab:skybkg}.}
\label{tab:tjump}
\end{center}
\end{table*}

No temperature jump is detected across the density discontinuity. However, the high temperatures of the ICM and the relatively short exposure time of the observations make the measurements too uncertain to exclude a temperature jump corresponding to a Mach number of $\sim 1.3$.

\subsubsection{X-ray morphological estimators}

In order to classify Abell 1033 as a possible merging cluster we analysed the X-ray surface brightness inside an aperture radius of $R_{ap}=500$ kpc (230\arcsec) using the methods described in \cite{Cassano2013}. 

We calculated the cluster emission centroid shift and the surface brightness concentration parameter. The centroid shift, $w$, is computed in a series of circular apertures centred on the cluster X-ray peak and is defined as the standard deviation of the projected separation between the peak and the centroid in unit of $R_{ap}$, as \citep{Poole2006,Maughan2008}:

\begin{equation}
w=\Big[\frac{1}{N-1}\sum (\Delta_i-\langle \Delta \rangle)^2\Big]^{1/2}\times \frac{1}{R_{ap}},
\label{Eq:w}
\end{equation}

\noindent where $\Delta_i$ is the distance between the X-ray peak and the centroid of the {\it i}th aperture.

The concentration parameter \citep{Santos2008}, $c$, is defined as the ratio of the peak over the ambient surface brightness, $S$, as:

\begin{equation}
c=\frac{S(r<100\,\mathrm{kpc})}{S(r<500\,\mathrm{kpc})}
\label{Eq:c}
\end{equation}

\target{} has a centroid shift $w=0.086 \pm 0.006$ and a concentration parameter $c=0.200 \pm 0.004$. These results put the cluster in an intermediate position being classified as disturbed for the centroid shift and as moderately relaxed for the concentration parameter \citep{Cassano2013}. \target{} is therefore different from clusters undergoing major mergers. This can be caused by the merger happening close to the line of sight (as suggested by the BCGs high differential velocity) or by the mass of the merging clusters being very different (minor merger). A third possibility is that the age of the merger is comparable to the sound crossing time of the X-ray gas (old merger). Thus the gas might have relaxed to some degree. This relaxation time can be less than the time it takes for the two subcluster dark matter halos and galaxies to reach their maximum separation and come back together.

\section{Discussion}
\label{sec:discussion}

The X-ray brightness distribution of \target{}, together with the high radial velocity of the BCGs, suggests a merger axis with a large line-of-sight component. The compression of the X-ray contours (see Fig.~\ref{fig:Chandra}) towards south shows that the merger axis must also have a north-south component, which is in line with the location of the subclusters detected by optical observations. The X-ray emission is mostly concentrated on the south sub-clusters. An explanation could be that the dense south core is being ram-pressured stripped as it moves through the ICM, while the north cluster did not enter this merger with a dense core, or perhaps its core was badly disturbed.

Extended radio emission (\textit{Source 1}, in Fig.~\ref{fig:sdss2}) with a steep spectrum ($\alpha \simeq -1.62\pm0.09$) and no optical counterpart is present close to the region where the X-ray surface brightness gradient is steepest. \textit{Chandra} data do not show any sign of a shock presence or temperature jump along what, at a first sight, could be interpreted as a radio relic. The extended radio source has some other peculiar characteristics compared to known radio relics: (i) the source is small (280 kpc), (ii) it has a high surface brightness, (iii) it is centrally-located in the cluster \citep[unexpected for radio relics;][]{Vazza2012} and (iv) its radiation is not polarized down to the 1\% level. Given these characteristics either the source is a very unusual radio relic or its nature is different.

\begin{figure}
\centering
\includegraphics[width=.5\textwidth]{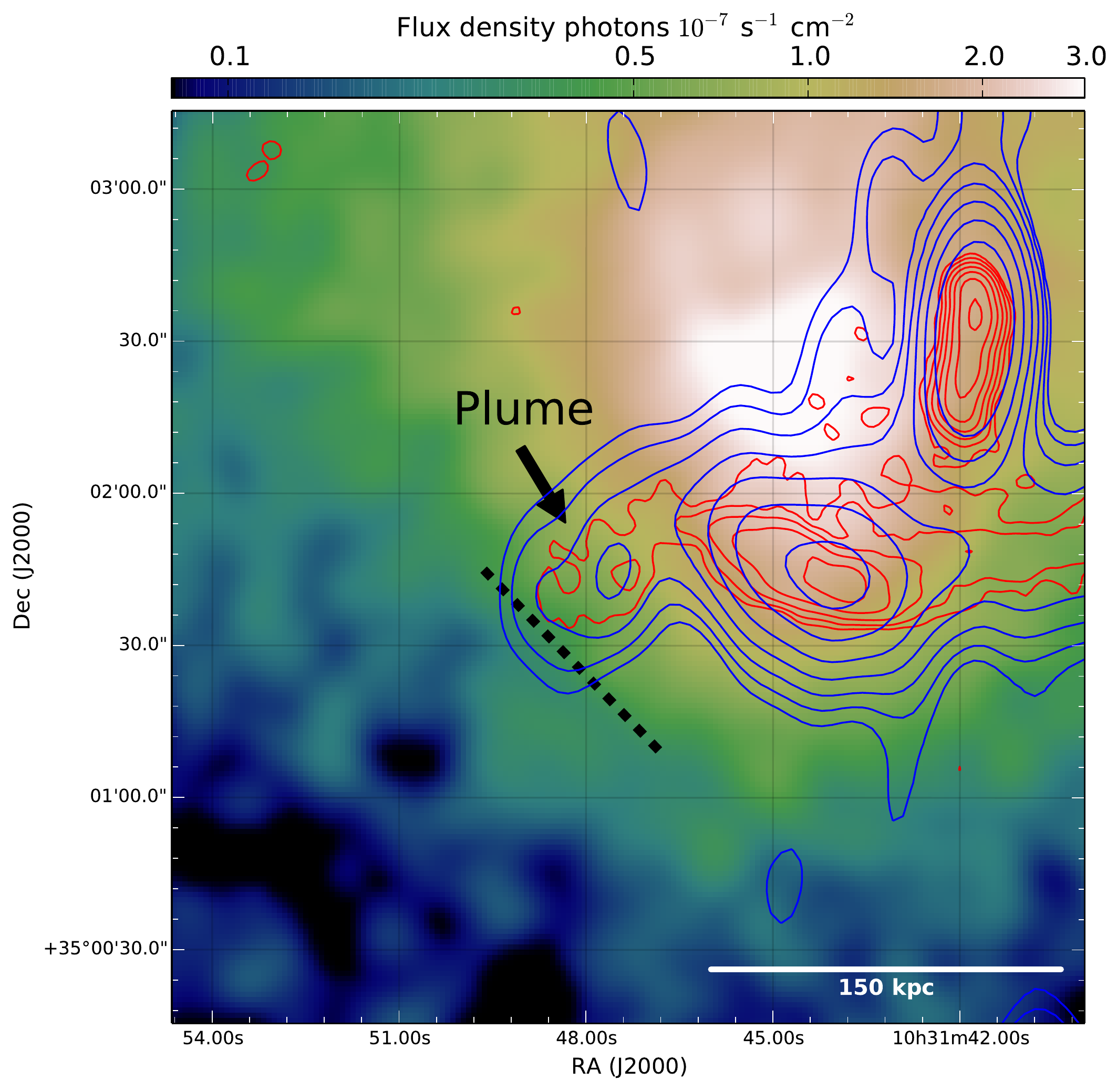}
\caption{Same as Fig.~\ref{fig:Chandra} but zoomed in on the density discontinuity and \textit{plume} region. The black dashed line shows the discontinuity location. Blue contours are from the low-resolution WSRT data (Fig.~\ref{fig:radio-wsrt}), while red contours are from the high-resolution VLA data (Fig.~\ref{fig:radio-vla}).}\label{fig:Chandra-zoom} 
\end{figure}

The radio morphology is in-line with the idea of \textit{Source 1} moving north to south. The radio brightness has a quite steep gradient and terminates sharply on the south edge, while the northern edge is disrupted and shows faint structures elongating towards the cluster centre. Interestingly, we discovered the presence of a density discontinuity in the south-west region of the cluster beyond the radius of the southern edge of \textit{Source 1}. The feature is located at the edge of a radio \textit{plume}, extending from the east side of the source (see Fig.~\ref{fig:Chandra-zoom}). The \textit{plume} could be therefore the result of a shock-driven \textit{re}-acceleration of CR electrons originated from \textit{Source 1} and points towards an association between \textit{Source 1} and the discontinuity. Reacceleration of CR electrons, in contrast with direct acceleration from the thermal pool, is thought to be required to power extended radio sources in galaxy clusters \citep[e.g.][]{Ensslin1998, Markevitch2005, Kang2011, Pinzke2013}.

\begin{figure}
\centering
\includegraphics[width=.5\textwidth]{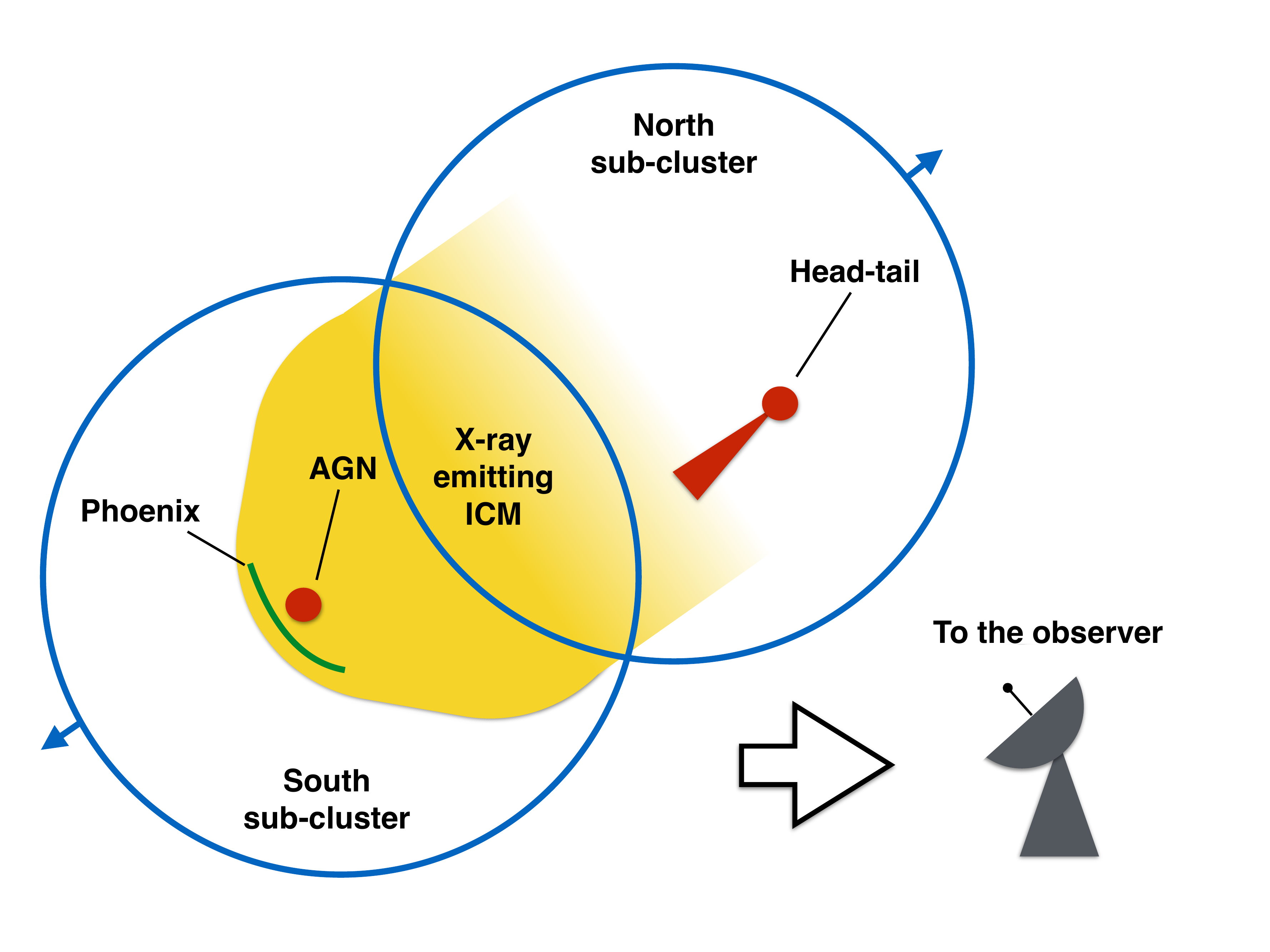}
\caption{Possible configuration of \target{} view from the side. The phoenix extensions is perpendicular to the line-of-sight. The observer view-point is on the right.}\label{fig:cartoon} 
\end{figure}

Given the difficulties to interpret \textit{Source 1} as a radio relic, we present an alternative explanation for the formation of this source which is related to the radio emission from another cluster member (SDSS J103143.69+350154.5, \textit{Source 4} in Fig.~\ref{fig:sdss4}), located just above (29 kpc in projection) the centre of the extended emission. The radio source, probably related to nuclear activity, is coincident with an early-type galaxy whose redshift ($z=0.125$) and proximity to the southern BCG suggest it is part of the southern sub-cluster (see Fig.~\ref{fig:cartoon}.). The extended radio emission has a brightness and a morphological configuration compatible with a Fanaroff-Riley I radio galaxy. Therefore, it could be a striking example of CRs produced by an old AGN outburst, displaced, and likely compressed by the galaxy cluster merger event through shock waves (i.e. a radio phoenix). While the thermal environment gets shocked, the radio plasma itself is only compressed adiabatically due to the much higher internal sound speed. Plasma compressed to become a radio phoenix close to the cluster centre is expected to produce bright emission with a very steep spectrum, as observed in this case \citep{Ensslin2001}. This scenario would explain the non-detection of a shock along its extension, the source location, morphology, spectral index, and brightness. High degrees of polarizations are expected as the magnetic field is aligned and compressed. The absence of polarization may be, in this case, an observational limit due to the poor resolution in combination with the strong Faraday depolarization due to the large quantity of intervening matter.

Another explanation could be that \textit{Source 1} traces the fading lobes of a radio galaxy (\textit{Source 4}) which became quiescent while moving north \citep[similar to the case of 3C~338;][]{Gentile2007}. However, the north and south edges of \textit{Source 1} show a movement of the lobes (compression south, rarefaction north) which is incompatible with old lobes left behind by an AGN moving north. Moreover, \textit{Source 1} would have had a narrow/wide angle tail morphology bended south, which is not seen in an environment where the other large radio galaxies show signs of interaction with the ICM.

Another possible explanation is that \textit{Source 1} was not displaced by a merging shock but by the bulk ICM motions induced by the merger event. In the simplest scenario, when the two sub-clusters merge, their dark matter and galaxy components pass each other relatively undisturbed, while the ICM interacts and slows down creating a relative velocity between the ICM and the galaxies. Assuming the sub-clusters just passed each other, as suggested by the presence of a brightness discontinuity and by the N$\rightarrow$S orientation of the head-tail which is in the north sub-cluster (see Fig.~\ref{fig:cartoon}.), we would expect the ICM in the south sub-cluster to be moving towards the north with respect to the cluster galaxies. Alternatively said, the member galaxies of the southern sub-cluster are moving south faster than the ICM is. This would have shifted the lobes towards the north rather than the south of the AGN, disfavouring this interpretation.

The most simple explanation that takes the above evidence into account is that the radio lobes have been displaced and compressed by a low Mach number merger shock moving outward in the ICM. The adiabatic compression boosts the lobes flux as in the scenario A (radio cocoon close to the cluster centre) in \cite{Ensslin2001}. Such a revived radio lobe cannot be very old ($<100$~Myr), thus must be close to the galaxy it originated from but displaced in the shock wave direction of propagation (i.e. south). Here we argue that \textit{Source 4} can be that source and that the merger has revived the radio cocoon. The detected discontinuity in the X-ray emission may well be (part of) the shock that originated the radio phoenix. Radio observations at low frequencies will provide a spectrum that can be used for deriving the age and the compression factor (subject to a number of assumptions) confirming this scenario.

Here we presented the simplest scenario which takes all the observable facts into account. Complications may rise if for example the ICM has relevant local velocities, \textit{Source 4} has an unusual peculiar velocity, projection effects mask the real shape of \textit{Source 1}, or the merger is more complex than a well-behaving dissociative merger (as a minor merger or a merger with a high impact parameter).

\section{Conclusions}
\label{sec:conclusions}
We presented \textit{Chandra}, WSRT, and VLA observations together with the analysis of SDSS images of the galaxy cluster \target{}. The main findings of the paper are:
\begin{itemize}
 \item The galaxy number density and redshift analysis performed on SDSS images shows a clear bimodal distribution. \target{} is composed of two subclusters of comparable richness.
 \item X-ray brightness maps show a disturbed cluster according to global morphological parameters. We report a steepening in the surface brightness profile towards south. \target{}'s merger axis is almost oriented along the line of sight. This projection effect could be responsible for the non-detection of merging shocks.
 \item The radio environment of \target{} is quite heterogeneous. We report the presence of a head-tail radio galaxy and of a wide-angle tail radio galaxy which are probably part of the northern sub-cluster. The N$\rightarrow$S orientation of the head-tail radio galaxy suggest a motion in line with the merger scenario.
 \item An extended, steep-spectrum ($\alpha \simeq 1.62\pm0.09$) radio source is reported. The source is likely a newborn radio phoenix originated by a merger-induced shock-waves which displaced (and possible compressed) the lobes of a radio galaxy located 29 kpc (in projection) above the source.
 \item We discovered a density discontinuity which could be caused by a weak shock with a Mach number of $1.28_{-0.11}^{+0.12}$, probably related to the recent merger. Interestingly, the discontinuity seems related not to the radio phoenix itself, but rather to a \textit{plume} extended from the radio phoenix (see Fig.~\ref{fig:Chandra-zoom}). We speculate that the discontinuity-related shock may have re-accelerated CR electrons present in the old radio galaxy lobe. On larger scales, this mechanism is thought to be responsible for generating giant radio halos. In some cases giant radio halos are in fact found to be spatially connected to a shock on their edge \citep[see e.g.][]{VanWeeren2012b, deGasperin2014c}.
\end{itemize}

Optical and multi-frequency radio observations of \target{} are planned and will shed light on this complex system. If the merger axis is slightly tilted N-S, then possible radio relics are expected to the north and/or to the south of the cluster. The picture will be also reinforced by the detection of a radio halo at low-frequencies.

\section*{Acknowledgements}
A.B. and M.B. acknowledge support by the research group FOR 1254 funded by the Deutsche Forschungsgemeinschaft: "Magnetisation of interstellar and intergalactic media:the prospects of low-frequency radio observations". G.A.O. acknowledges support by NASA through a Hubble Fellowship grant HST-HF2-51345.001-A awarded by the Space Telescope Science Institute, which is operated by the Association of Universities for Research in Astronomy, Incorporated, under NASA contract NAS5-26555. R.J.W. is supported by NASA through the Einstein Postdoctoral grant number PF2-130104 awarded by the Chandra X-ray Center, which is operated by the Smithsonian Astrophysical Observatory for NASA under contract NAS8-03060. Part of this work performed under the auspices of the U.S. DOE by LLNL under Contract DE-AC52-07NA27344.

Funding for SDSS-III has been provided by the Alfred P. Sloan Foundation, the Participating Institutions, the National Science Foundation, and the U.S. Department of Energy Office of Science. The SDSS-III web site is \url{http://www.sdss3.org}.


This research has made use of the NASA/IPAC Extragalactic Database (NED) which is operated by the Jet Propulsion Laboratory, California Institute of Technology, under contract with the National Aeronautics and Space Administration.

This research has made use of NASA's Astrophysics Data System.

\bibliographystyle{mn2e}
\bibliography{abell1033}
\bsp

\label{lastpage}

\end{document}